\documentclass[preprint,showpacs,preprintnumbers,amsmath,amssymb]{revtex4}

\usepackage{graphicx}
\usepackage{dcolumn}
\usepackage{bm}
\usepackage{multirow}
\usepackage{amsmath}

\begin{document}
\title{{\it Ab initio} thermodynamic study of SnO$_2$(110) surface in an O$_2$ and NO environment: a fundamental understanding of gas sensing mechanism for NO and NO$_2$}

\author{Chol-Jun Yu,$^1$\footnote{Chol-Jun Yu: ryongnam14@yahoo.com} Yun-Hyok Kye,$^1$ Song-Nam Hong,$^1$ Un-Gi Jong,$^1$ Gum-Chol Ri,$^1$ Chang-Song Choe,$^2$ Kwang-Hui Kim$^2$ and Ju-Myong Han$^2$}
\affiliation{$^1$Department of Computational Materials Design (CMD) and $^2$Department of Semiconducting Materials, Faculty of Materials Science, Kim Il Sung University, Ryongnam-Dong, Taesong District, Pyongyang, DPR Korea}

\date{\today}

\begin{abstract}
For the purpose of elucidating the gas sensing mechanism of SnO$_2$ for NO and NO$_2$ gases, we calculate the phase diagram of SnO$_2$(110) surface in contact with an O$_2$ and NO gas environment by means of {\it ab initio} thermodynamic method. Firstly we build a range of surface slab models of oxygen pre-adsorbed SnO$_2$(110) surfaces using (1$\times$1) and (2$\times$1) surface unit cells and calculate their Gibbs free energies considering only oxygen chemical potential. The fully reduced surface containing the bridging and in-plane oxygen vacancies in the oxygen-poor condition, while the fully oxidized surface containing the bridging oxygen and oxygen dimer in the oxygen-rich condition, and the stoichiometric surface in between, were proved to be most stable. Using the selected plausible NO-adsorbed surfaces, we then determine the surface phase diagram of SnO$_2$(110) surfaces in ($\Delta\mu_\text{O}$, $\Delta\mu_\text{NO}$) space. In the NO-rich condition, the most stable surfaces were those formed by NO adsorption on the most stable surfaces in contact with only oxygen gas. Through the analysis of electronic charge transferring and density of states during NO$_x$ adsorption on the surface, we provide a meaningful understanding about the gas sensing mechanism.
\end{abstract}

\maketitle

\section{\label{sec_intro}Introduction}
Tin dioxide (SnO$_2$) is a wide band gap ($E_g=3.6$ eV) $n$-type semiconducting oxide that has received much attention for the past decades due to its remarkable technological applications such as transparent electrodes in solar cells~\cite{Wagner} and catalytic supporting materials~\cite{Park}. In particular, it is widely used as solid state chemical sensors to both oxidizing (e.g., CO$_2$, NO$_2$) and reducing (CO, NO) gases~\cite{Hahn,Zeng,Maiti,Leblanc,Canevali,Anjali,Epifani06,Epifani08,Liu,Wang16}. In all these applications, the key for governing device functionality is the properties of SnO$_2$ surface or its interface with functional organic molecules. Moreover, the optical, electronic and catalytic properties of SnO$_2$ depend critically on surface modifications such as impurities, defects or adsorbate~\cite{Zhu,Wei,Jin,Korber}, and especially, its electric conductivity varies sensitively upon adsorption of gas molecule at the surface~\cite{Batzill05,Wang1,Wang2}. Such conductance change of the sensing layer was well established to be the basic detection principle of the chemical gas sensor, and yet a fundamental understanding of the key phenomena at the SnO$_2$ surface remains debatable~\cite{Gopel,Ippommatsu,Cassia,Batzill,Das}. This gap in our understanding is also a problem for tailoring efficient gas sensors based on SnO$_2$ with desirable sensing characteristics, such as high sensitivity and selectivity, long-term stability, and fast response time.

In this work, we address this knowledge gap, by focusing on the composition and structure of the SnO$_2$(110) surface in an O$_2$ and NO environment with {\it ab initio} atomistic thermodynamics to give a concise microscopic view of NO$_x$ (NO and NO$_2$) gas sensing by SnO$_2$. NO and NO$_2$ gases, byproducts of running car and industry, are one of the most toxic air pollutants and the main source of acid rain~\cite{Anjali,Liu}. Moreover, when one is over exposed (lower tolerance limit $\sim$5 ppm), it may lead to pulmonary disease and even the loss of human life in extreme cases. In recent years, therefore, the significant importance of NO$_x$ sensing has been emphasized for protecting human health and environment.

Using density functional theory (DFT) calculations, it was found that oxygen vacancies ($V_\text{O}$) are the main cause of the (unintentional) $n$-type conductivity by forming shallow donor levels at the bottom of the conduction band with the mobility of electrons from Sn(II) to Sn(IV) sites in bulk SnO$_2$~\cite{Godinho}, though other point defects such as Sn interstitial (Sn$_i$) and substitutional hydrogen impurities (H$_\text{O}$) play a certain role in the growth and processing environment~\cite{Kilic,Singh}. This prediction was proved to be consistent with the experiment~\cite{Samson,Maier}, and extended from bulk vacancy to surface oxygen vacancies at the SnO$_2$ surface~\cite{Trani,Habgood,Yamaguchi,Guo,Maki,Duan,Cox}.

The stoichiometric SnO$_2$(110) surface, which is accepted to be the most stable among the crystal faces, consists of O-(Sn$_2$O$_2$)-O layers. Here, two-coordinated oxygen atom in the outermost layer is referred to as bridging oxygen atom. The removal of bridging oxygen atom, oxygen atom in the last Sn$_2$O$_2$ plane, and oxygen atom in the next plane, leads to creation of a {\it bridging vacancy}, an {\it in-plane vacancy}, and a {\it sub-bridging vacancy}, respectively~\cite{Habgood,Batzill05}. In the removal of bridging oxygen atoms, two electrons are left, resulting in the reduction of tin from six-coordinated Sn$^{4+}$ ion to four-coordinated Sn$^{2+}$ ion ({\it i.e.}, becoming the reduced surface), with which an oxygen molecule can interact.

\begin{figure}[!th]
\begin{center}
\includegraphics[clip=true,scale=0.5]{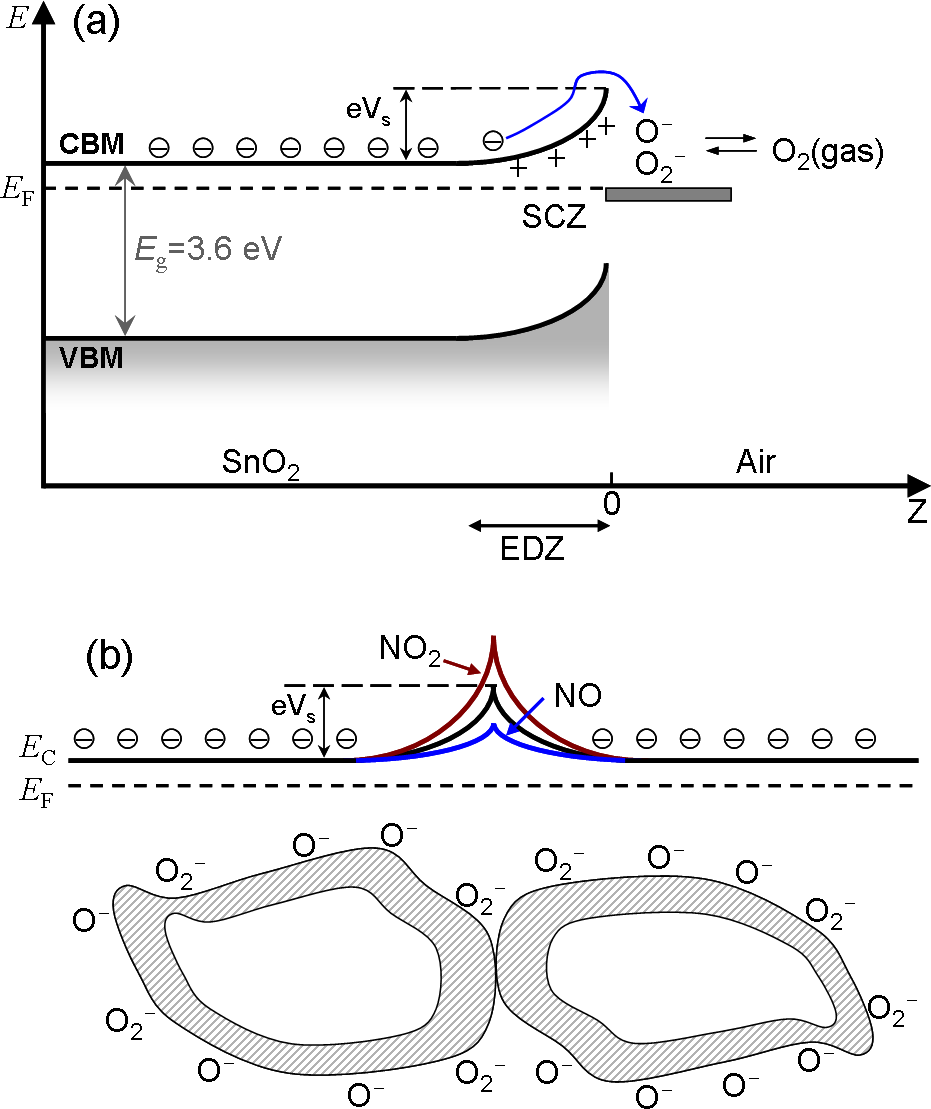}
\end{center}
\caption{\label{fig_mecha}(Color online) (a) Schematic view of energy band bending of SnO$_2$ surface by donating electrons to surface oxygen molecules over the Schottky barrier ($eV_s$), leading to the creation of electron depletion zone (EDZ) and positive space charge zone (SCZ). (b) Schematic change of Schottky barrier at a grain contact when adsorbing NO or NO$_2$ molecule.}
\end{figure}

The pre-adsorption of oxygen on the reduced SnO$_2$ surface is the first stage of the reducing gas sensing action, since sensors are in general exposed to air atmosphere before the introduction of gas~\cite{Trani,Habgood,Yamaguchi}. Oxygen ionosorbs onto the surface, trapping conduction electrons from SnO$_2$ and creating a superoxo O$_2^-$ or an atomic O$^-$ ion at the surface~\cite{Shen,Pang}. This happens because the lowest unoccupied molecular orbitals of the adsorbate lie below the Fermi level $E_\text{F}$ of the solid~\cite{Madou}. As a consequence, the surface has a net negative charge causing an electric field, which induces upward surface band bending, resulting in the push of the Fermi level into the band gap of the solid, the reduce of the charge carrier concentration and thus the creation of the electron depletion zone (EDZ)~\cite{Batzill}. Depleting electrons leads to a generation of positive space charge zone (SCZ) that compensates for the negative surface charge, and therefore, alters the sheet conductance of the surface layer~\cite{Madou}. Furthermore, the band bending at the surface has an additional effect in the case of polycrystalline phase, i.e., the formation of Schottky barriers ($eV_s$) at grain boundaries across which conduction electrons have to overcome to carry the current. When gas molecule is introduced into the pre-adsorbed SnO$_2$ surface, there is an interaction between the molecule and the surface oxygen species O$^-$ and/or O$_2^-$. This causes a reverse (rise) of the band bending and thus a decrease (increase) of the barrier at grain boundaries, resulting in an increased (decreased) conductivity for reducing (oxidizing) gases~\cite{Batzill} (see Fig.~\ref{fig_mecha}).

There exist several experimental works demonstrating NO~\cite{Canevali,Leblanc} as well as NO$_2$~\cite{Leblanc,Anjali,Liu,Epifani08,Epifani06} sensing of SnO$_2$ in the form of nanocrystals and thick porous films. A significant decrease of surface resistance was observed when introducing NO gas in air, due to the injection of electrons from NO to oxide with formation of oxygen vacancies~\cite{Canevali}. On the contrary, NO$_2$ interact with SnO$_2$ by trapping the free electrons both directly and indirectly through the ionosorbed surface oxygen species, increasing the potential barrier across grain boundaries and hence increasing its resistance~\cite{Liu}. Epifani et al.,~\cite{Epifani08} emphasized the role of surface oxygen vacancies in the NO$_2$ sensing properties of SnO$_2$ nanocrystals by performing spectroscopic measurements and DFT calculations of the NO$_2$/SnO$_2$ system. They suggested that the interaction between NO$_2$ and the surface occurs through the oxygen vacancy sites, and that the presence of bridging oxygen vacancies strongly enhances the charge transfer from the surface to NO$_2$. Similar finding was obtained for NO$_2$ sensing by WO$_3$ nanowires~\cite{Qin}. Prades et al.,~\cite{Prades,Prades07} revealed using DFT calculations that NO and NO$_2$ molecules are adsorbed on bridging oxygen sites and bridging oxygen vacancy sites at the SnO$_2$(110) surface. Compared to the CO sensing mechanism of SnO$_2$(110) surface~\cite{Wang,Yang,Shan,Ciriaco,Duere}, the sensing properties strongly depends upon the concentration of oxygen in the ambient atmosphere: CO reacts with either directly stoichiometric surface or oxygen species O$_2^-$ or O$^-$, accompanying the release of electrons to the surface with or without formation of CO$_2$~\cite{Wang}.

However, the preceding NO$_x$/SnO$_2$(110) studies have either focused on surface adsorption without consideration of surrounding gas effect~\cite{Prades,Prades07}, or used only oxygen chemical potential to get the energetically lowest surface in contact with oxygen gas~\cite{Epifani08}. Therefore, these studies could not investigate the effect of ambient NO gas on NO$_x$ gas sensing of SnO$_2$-based gas sensor, and the suggested mechanisms could be limited. Many aspects of the sensing mechanism including the adsorption sites and charge transfer are not yet fully understandable. In the present study, we aim to clarify the details of O$_2$ and NO$_x$ adsorptions on SnO$_2$(110) surfaces, and to calculate the surface phase diagram of SnO$_2$(110) surface in contact with O$_2$ and NO gases, using {\it ab initio} DFT calculations with the inclusion of van der Waals interaction.

\section{\label{sec_method}Method}

\subsection{\label{subsec_thermo}{\it Ab initio} thermodynamics for SnO$_2$ surface in O$_2$ and NO environment}
In this work, we consider the thermodynamic stability of the SnO$_2$ surface in contact with O$_2$ and NO gas reservoirs. In the equilibrium state with such environmental gases, the most stable surface at temperature $T$ and partial pressures $\{p_{\text{O}_2}, p_{\text{NO}}\}$ is one to minimize the surface free energy given as
\begin{equation}
\label{surf_ene1}
\gamma(T, p_{\text{O}_2}, p_\text{NO})=\frac{1}{2A}[G_\text{slab}-N_\text{Sn}\mu_\text{Sn}(T) -N_\text{O}\mu_\text{O}(T, p_{\text{O}_2})-N_\text{NO}\mu_\text{NO}(T, p_\text{NO})],
\end{equation}
where $G_\text{slab}$ is the Gibbs free energy of the slab with two equivalent surface area A, and $\mu_i$ ($i$=Sn, O, NO in this work) is the chemical potential of the species $i$ with its number of atoms or molecules $N_i$ contained in the slab system~\cite{Reuter03,Reuter03b,Rogal}. For the NO-adsorbed SnO$_2$(110) surfaces, it is acceptable without doubt that Sn is in bulk phase, and O$_2$ and NO are in gaseous phases.

With respect to the chemical potentials of the species, we should consider various thermodynamic constraints. For example, the chemical potentials of Sn and O$_2$ are related with each other through the Gibbs free energy per formula unit of bulk SnO$_2$, $\text{g}_{\text{SnO}_2}$, based on the fact that bulk SnO$_2$ material is in equilibrium with the O$_2$ gas reservoir at not too low temperature. That is,
\begin{equation}
\label{chem_pot1}
 \mu_\text{Sn}+\mu_{\text{O}_2}=\text{g}_{\text{SnO}_2}.
\end{equation}
Considering the equation $\mu_{\text{O}_2}=2\mu_\text{O}$ and replacing $\mu_\text{Sn}$ in Eq.~\ref{surf_ene1} with Eq.~\ref{chem_pot1}, the surface free energy can be rewritten in a way of eliminating its dependence on $\mu_\text{Sn}$ like,
\begin{equation}
\label{surf_ene2}
\gamma(T, p_{\text{O}_2},p_\text{NO})=\frac{1}{2A}[G_\text{slab}-N_\text{Sn}\text{g}_{\text{SnO}_2} 
+(2N_\text{Sn}-N_\text{O})\mu_\text{O}(T, p_{\text{O}_2})-N_\text{NO}\mu_\text{NO}(T, p_\text{NO})].
\end{equation}
The chemical potentials of O and NO are also constrained by the thermodynamic equilibrium condition with the surrounding gas reservoirs, and can be rewritten using the DFT total energies $E$ at 0 K as follows,
\begin{flalign}
\label{chem_pot21}
\mu_{\text{O}}(T, p_{\text{O}_2})&=E_\text{O}+\Delta\mu_\text{O}(T, p_{\text{O}_2}),\\
\label{chem_pot22}
\mu_{\text{NO}}(T, p_{\text{NO}})&=E_\text{NO}+\Delta\mu_\text{NO}(T, p_\text{NO}),
\end{flalign}
where $E_{\text{O}_2}(=2E_\text{O})$ and $E_\text{NO}$ are the total energies of isolated O$_2$ and NO molecule, respectively, and $\Delta\mu$ is the difference between the Gibbs free energy and DFT total energy. Inserting Eq.~\ref{chem_pot21} and Eq.~\ref{chem_pot22} into Eq.~\ref{surf_ene2} results in the following equation,
\begin{multline}
\label{surf_ene3}
\gamma(T, p_{\text{O}_2}, p_\text{NO})
=\frac{1}{2A}[G_\text{slab}-N_\text{Sn}\text{g}_{\text{SnO}_2}+(2N_\text{Sn}-N_\text{O})E_\text{O}-N_\text{NO}E_\text{NO}] \\
+\frac{1}{2A}[(2N_\text{Sn}-N_\text{O})\Delta\mu_\text{O}\left(T, p_{\text{O}_2}\right)-N_\text{NO}\Delta\mu_\text{NO}(T, p_\text{NO})].
\end{multline}
Moreover, the difference of the Gibbs free energies of the surface slab and bulk unit cell can be approximated by the difference of the corresponding DFT total energies with a good reason that for SnO$_2$ the vibrational and entropical contributions to that are negligible and cancel each other, as other transition metal oxide materials~\cite{Reuter03b,Rogal}. Therefore, defining the NO-adsorbed surface formation energy with only DFT total energies as
\begin{equation}
\label{ene_surf}
\gamma^f(0, 0)=\frac{1}{2A}[E_\text{slab}-N_\text{Sn}E_{\text{SnO}_2}  +(2N_\text{Sn}-N_\text{O})E_\text{O}-N_\text{NO}E_\text{NO}],
\end{equation}
the surface free energy can be approximately rewritten as follows,
\begin{equation}
\label{surf_ene4}
\gamma(T, p_{\text{O}_2}, p_\text{NO})\approx \gamma^f(0, 0) 
+\frac{1}{2A}[(2N_\text{Sn}-N_\text{O})\Delta\mu_\text{O}(T, p_{\text{O}_2}) -N_\text{NO}\Delta\mu_\text{NO}(T, p_\text{NO})].
\end{equation}

In Eq.~\ref{surf_ene4}, the surface free energy at a certain temperature and partial pressures is expressed as a linear function of the chemical potentials of environmental gases. Therefore, it is useful to consider the relating ranges of the chemical potentials. Due to the constraint that the bulk phase of SnO$_2$ in equilibrium state remains stable, $\Delta\mu_{\text{O}}(T, p_{\text{O}_2})$ has the lower limit ({\it i.e.,} O$_{\text{poor}}$ limit) of $\Delta G^f_{\text{SnO}_2}(0, 0)=E_{\text{SnO}_2}-E_\text{Sn}-E_{\text{O}_2}$, which is the formation energy of the bulk SnO$_2$ ($-6.087$ eV in Table.~\ref{tab_xccheck}). On the other hand, its upper limit ({\it i.e.,} O$_\text{rich}$ limit) is 0 from the fact that DFT total energy is the maximum value for the chemical potential of oxygen molecule. Therefore, the following inequality for $\Delta\mu_\text{O}$ is established,
\begin{equation}
\label{delta_uo}
\frac{1}{2}{\Delta}G^f_{\text{SnO}_2}(0,0)<\Delta\mu_{\text{O}}(T, p_{\text{O}_2})<0.
\end{equation}
With respect to the thermodynamic constraint for the chemical potential of NO molecule, we ignore the possibility of the gas phase reaction NO+1/2O$_2\rightarrow$NO$_2$ because of a high energy barrier for that reaction. However, the situation becomes different when the SnO$_2$ surface is added: NO can be readily oxidized by a catalysis of SnO$_2$ surface. To prevent the reduction of bulk SnO$_2$ in a pure NO environment, therefore, the following inequality should be satisfied,
\begin{equation}
\label{delta_uno1}
\mu_{\text{SnO}_2}+2\mu_{\text{NO}}<\mu_{\text{Sn}}+2\mu_{\text{NO}_2}.
\end{equation}
With this inequality and similar argument to oxygen for the upper limit, the range for $\mu_{\text{NO}}$ variation is as follows,
\small
\begin{equation}
\label{delta_uno2}
\begin{cases}
\Delta\mu_{\text{NO}}(T, p_{\text{NO}})<\Delta\mu_{\text{O}}(T, p_{\text{O}_2}) -\Delta{G^f_{\text{SnO}_2}}(0,0)+\Delta{E_{\text{mol}}} \\ 
\Delta\mu_{\text{NO}}(T, p_{\text{NO}})<0,
\end{cases}
\end{equation}
\normalsize
\begin{equation}
\label{delta_emol}
\Delta E_\text{mol} = E_{\text{NO}_2}^\text{bind}-E_{\text{NO}}^\text{bind} - \frac{1}{2}E_{\text{O}_2}^\text{bind}.
\end{equation}
In our calculation, $\Delta E_\text{mol}=-1.62$ eV and $\Delta G^f_{\text{SnO}_2}(0,0)=-6.087$ eV (these are $-0.56$ eV and $-6.053$ eV in experiment~\cite{Lide}), and therefore, we should rely on the second inequality in Eq.~\ref{delta_uno2}, which is $\Delta\mu_{\text{NO}}(T, p_\text{NO})<0$.

The variation of the chemical potentials of oxygen and NO can be described using the pressure in the reference state (often atmosphere pressure $p^\circ$)~\cite{Rogal} like
\begin{flalign}
\label{deltchem1}
\Delta\mu_\text{O}(T, p_{\text{O}_2})&=\Delta\mu_\text{O}(T, p^\circ)+\frac{k_BT}{2}\ln\frac{p_{\text{O}_2}}{p^\circ},\\
\label{deltchem2}
\Delta\mu_\text{NO}(T, p_\text{NO})&=\Delta\mu_\text{NO}(T, p^\circ)+k_BT\ln\frac{p_\text{NO}}{p^\circ},
\end{flalign}
where $p^\circ=10^5$ Pa and $k_B$ the Boltzmann constant. We referred to the experimental values for $\Delta\mu_\text{O}(T, p^\circ)$ and $\Delta\mu_\text{NO}(T, p^\circ)$~\cite{Lide}.

\subsection{\label{subsec_model}Computational method}
All calculations in this work were performed by using the projector augmented wave (PAW) method as implemented in the Quantum ESPRESSO package~\cite{QE}. The PAW potentials, where the valence electron configurations are 4d$^{10}$5s$^2$5p$^2$ for Sn, 2s$^2$2p$^4$ for O, and 2s$^2$2p$^3$ for N atoms respectively, were used as provided in the code~\footnote{We used the PAW potentials Sn.pbesol-dn-kjpaw\_psl.0.2.UPF, O.pbesol-dn-kjpaw\_psl.0.2.UPF, and N.pbesol-dn-kjpaw\_psl.0.2.UPF from http://www.quantum-espresso.org.}. The major computational parameters were chosen as the plane wave cut-off energy to be 50 Ry and $k$-point meshes to be (6$\times$6$\times$4) and (8$\times$4$\times$1) for bulk and surface. With these parameters, the total energy of bulk and the surface formation energy were converged within 1 meV per bulk atom and 0.005 J/m$^2$, respectively. All the atomic positions were fully relaxed until the forces on each atom were less than 0.01 eV/\AA.

\begin{figure}[!th]
\begin{center}
\includegraphics[clip=true,scale=0.35]{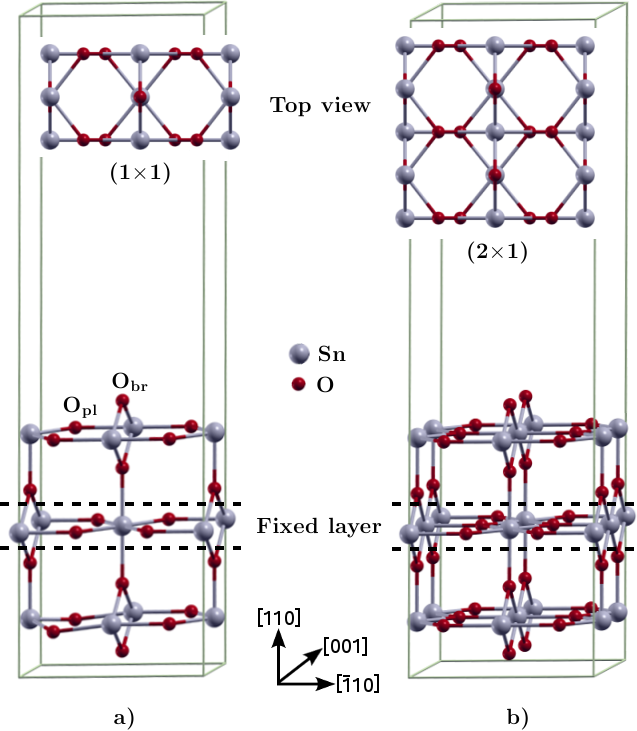}
\end{center}
\caption{\label{fig_slab}(Color online) Ball-and-stick model of supercells for stoichiometric SnO$_2$(110) surfaces with a) (1$\times$1) and b) (2$\times$1) surface unit cells, showing locations of the two distinct surface oxygens, O$_{\text{br}}$ for bridging oxygen and O$_{\text{pl}}$ for in-plane oxygen. Supercells consist of three O-(Sn$_2$O$_2$)-O atomic trilayers, where (Sn$_2$O$_2$) layers in the middle indicated by dashed lines are fixed, and vacuum layers of 12 \AA~thickness.}
\end{figure}

To allow 3D periodic simulations of surface, we have built supercells using inversion-symmetric slabs and (2$\times$1) as well as (1$\times$1) surface unit cells, of which lattice parameters are those of the bulk determined in this work. The supercells consisted of three O-(Sn$_2$O$_2$)-O atomic trilayers and vacuum layers of 12 \AA, as illustrated in Fig.~\ref{fig_slab} and used also in the previous DFT work~\cite{Wang}. When increasing the atomic layers from three to five trilayers, the surface energy increases by only 0.01 J/m$^2$, which is within the numerical noise. As mentioned in the introduction, there are two distinct surface oxygens, denoted as O$_{\text{br}}$ for bridging oxygen and O$_{\text{pl}}$ for in-plane oxygen, and the reduced and subreduced surface models could be built by removing the O$_{\text{br}}$ and O$_{\text{pl}}$ atoms properly from the stoichiometric surface. And simple cubic supercells with a lattice constant of 12 \AA~were used to make modeling of free molecules with the usage of only $\Gamma$ point in the reciprocal space.

We have tested the various exchange-correlation (XC) functionals; Perdew-Zunger local density approximation (PZ-LDA)~\cite{PZlda}, Perdew-Burke-Ernzerhof generalized gradient approximation (PBE-GGA)~\cite{PBE}, and PBE revised version for solid (PBEsol)~\cite{PBEsol}. In addition, the van der Waals (vdW) dispersive energy correction to the PBEsol energy, which is expected to play an important role in the surface adsorption, was also considered by using either Grimme method~\cite{Grimme} or exchange-hole dipole-moment (XDM) method~\cite{XDM}. In Table~\ref{tab_xccheck}, we list the calculated lattice constants, formation energies of bulk SnO$_2$, surface formation energy of (110) stoichiometric surface, and N-O bond length in NO molecule, according to the different XC functionals.
\begin{table}[!ht]
\begin{center}
\small
\caption{\label{tab_xccheck}Calculated lattice constants ($a$ and $c$), formation energy ($\Delta G^f$) of bulk SnO$_2$ with a rutile structure, surface formation energy ($\gamma^f$) of stoichiometric (110) surface, and N-O bond length in NO molecule ($d_\text{N-O}$) using different XC functionals.}
\begin{tabular}{lcclcc}
\hline
& $a$ & $c$ & ~~$\Delta G^f$ & $\gamma^f$ & $d_\text{N-O}$ \\
\cline{2-3}
XC & \multicolumn{2}{c}{(\AA)} & ~~(eV) & (J/m$^2$) & (\AA) \\
\hline
 PZ-LDA        & 4.722 & 3.204 & $-6.405$ & 1.394 & 1.150 \\
 PBE-GGA       & 4.800 & 3.243 & $-5.421$ & 1.003 & 1.160 \\
 PBEsol        & 4.752 & 3.216 & $-5.797$ & 1.189 & 1.156 \\
 PBEsol+Grimme & 4.740 & 3.231 & $-6.314$ & 1.679 & 1.156 \\
 PBEsol+XDM    & 4.750 & 3.218 & $-6.087$ & 1.487 & 1.156 \\
\hline
 Reference & ~4.737$^a$ & ~3.186$^a$ & $-6.070^b$ & ~1.210$^c$ & ~1.151$^b$ \\
\hline
\end{tabular} \\
$^a$Experiment~\cite{Haines} \\ 
$^b$Experiment~\cite{Lide} \\ 
$^c$FP-LAPW calculation~\cite{Batzill05}
\end{center}
\end{table}
\normalsize

With respect to the bulk properties of SnO$_2$, inherent overestimation of binding by LDA versus underestimation by GGA was also observed, while PBEsol improved the accuracy over GGA. In particular, the calculated values for the bulk formation energy largely deviate for LDA ($-6.41$ eV/f.u.) as well as for GGA ($-5.42$ eV/f.u.) when compared with the experiment ($-6.07$ eV/f.u.)~\cite{Lide}. We have chosen the PBEsol+XDM functional, since it can reproduce well the bulk formation energy ($-6.09$ eV/f.u.) and lattice constants from experiment. It was found that the calculated bond lengths with PBEsol+XDM were 1.210 \AA~in O$_2$ and 1.156 \AA~in NO, being in good agreement with the experimental values. In addition, for the case of free NO$_2$ molecule, N-O bond length and O-N-O bond angle were calculated to be 1.200 \AA~and 133.8$^{\circ}$, which are also consistent with the experimentally observed values of 1.190 \AA~and 134.1$^{\circ}$~\cite{Lide}. We note that the calculated band gap (1.2 eV) of bulk and binding energies of free NO ($-11.71$ eV), NO$_2$ ($-17.63$ eV) and O$_2$ ($-8.60$ eV) were deviated far from the corresponding experimental values (3.6, $-6.53$, $-9.70$, $-5.16$ eV)~\cite{Lide}, but these inaccurate values do not affect critically on the accuracy of surface related property calculation that will be carried out in this work.

\section{\label{sec_result}Result and discussion}
\subsection{\label{subsec_o2surf}Surface phase diagram in O$_2$ environment}
Before addressing the NO$_x$ adsorption onto SnO$_2$(110) surface, we first consider the phase diagram of the SnO$_2$(110) surfaces in the O$_2$ environment. It is worthy pointing out that the stoichiometric SnO$_2$(110) surface is type 2 following the Tasker's classification scheme~\cite{Tasker}, meaning no net dipole moment perpendicular to the surface consisted of symmetric trilayers, and thus it can exist as stable configuration without any reconstruction in vacuum. However, when the surface is brought into contact with air, a variety of surface reconstruction or stoichiometry variation could be occurred due to the interaction with mostly oxygen gas~\cite{Reuter03}. Depending on the preparation condition and sample history as well as temperature and oxygen pressure, $c(2\times2)$, $(4\times1)$, $(2\times1)$, $(1\times2)$ reconstructions, and $(1\times1)$ stoichiometric variation were identified by numerous experiments including low energy electron diffraction (LEED) and atomically resolved scanning tunneling microscopy (STM)~\cite{Batzill,Pang,Fresart,Atrei,Cox}. Based on these experimental observations, therefore, it is natural to regard that the stoichiometric SnO$_2$(110) surface at certain temperature and pressure given by the chemical potential of oxygen is expected to be not only reduced (oxygen depletion) but also oxidized (oxygen adsorption). This is recognized as a preliminary stage of gas sensing~\cite{Epifani08,Wang,Habgood,Yang,Trani}, since it has become clear that pre-adsorbed oxygen species O$_2^-$ and O$^-$ on the surface grab electrons from the surface, resulting in an increase of the surface resistance~\cite{Hahn,Wang16,Epifani08,Canevali}.

\begin{figure}[!th]
\begin{center}
\includegraphics[clip=true,scale=0.26]{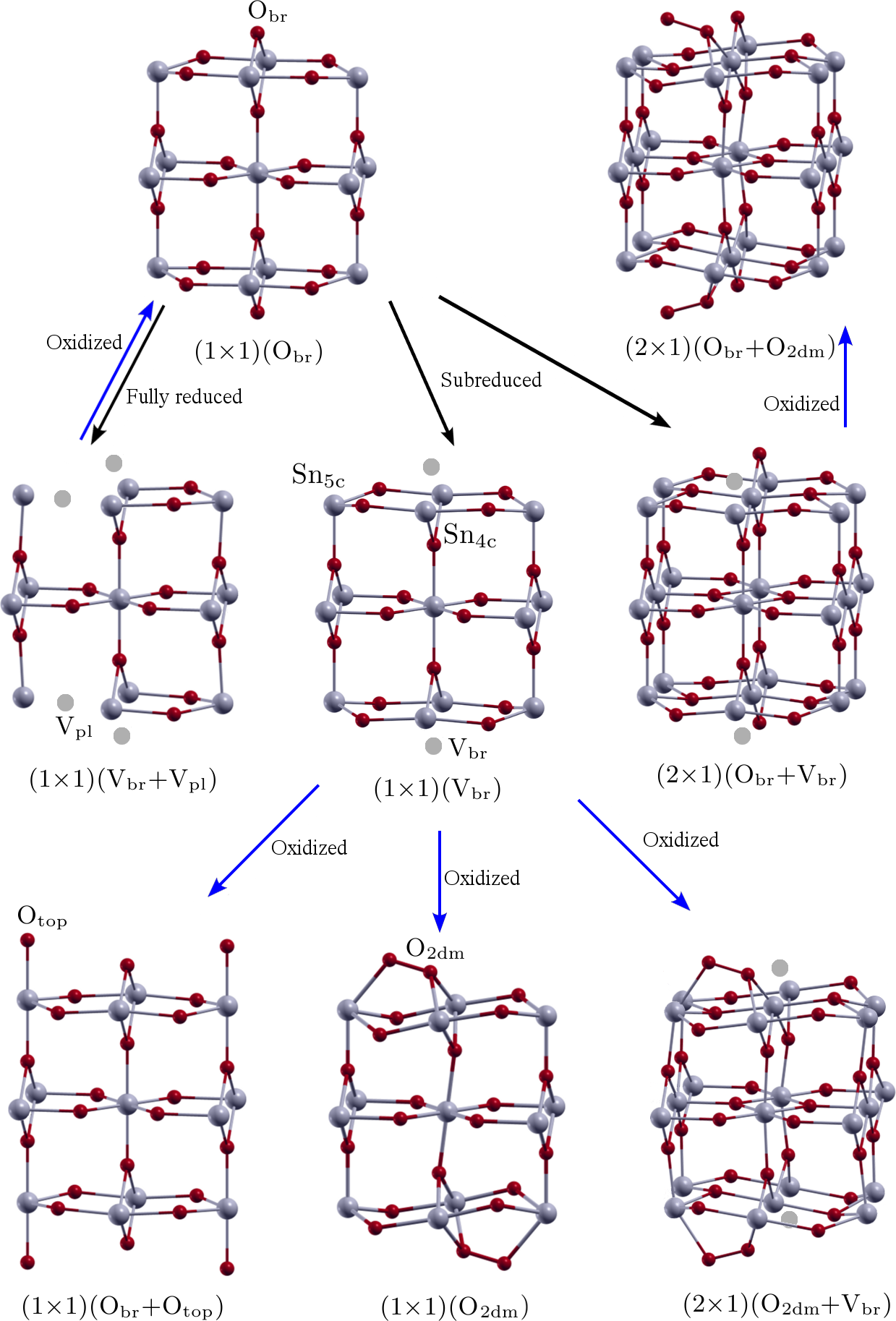}
\end{center}
\caption{\label{fig_surf1}(Color online) SnO$_2$(110) surface models in an O$_2$ environment. O$_\text{br}$, O$_\text{top}$ and O$_\text{2dm}$ represent the bridging oxygen, top oxygen over the five-coordinated Sn atom denoted Sn$_\text{5c}$ and oxygen dimer formed over the surface between Sn$_\text{5c}$ and the four-coordinated Sn (Sn$_\text{4c}$) atoms, and V$_\text{br}$ and V$_\text{pl}$ indicate the bridging and in-plane oxygen vacancies.}
\end{figure}
In this work we suggest eight different models for SnO$_2$(110) surface in contact with oxygen gas using (1$\times$1) and (2$\times$1) surface unit cells, as shown in Fig.~\ref{fig_surf1}. For these models we introduce a notation like (surface unit cell)(pre-adsorbed oxygen + oxygen vacancy). The starting point is the stoichiometric surface denoted as (1$\times$1)(O$_\text{br}$) (no presence of oxygen vacancy). When removing both bridging and in-plane oxygens from the stoichiometric surface, the fully reduced surface denoted as (1$\times$1)(V$_\text{br}$+V$_\text{pl}$) is generated. The removal of either all or half the bridging oxygens leads to the subreduced surfaces, referred to as (1$\times$1)(V$_\text{br}$) or (2$\times$1)(O$_\text{br}$+V$_\text{pl}$). These three models are served as the substrates for oxygen adsorption as in the previous DFT works~\cite{Habgood,Trani}. We note that the stoichiometric surface (1$\times$1)(O$_\text{br}$) is also recognized as the ``oxidized'' surface from (1$\times$1)(V$_\text{br}$+V$_\text{pl}$) surface~\cite{Habgood}. When an oxygen molecule approaches the subreduced (1$\times$1)(V$_\text{br}$) surface, it can be adsorbed onto the surface in either dissociative way -- leading to (1$\times$1)(O$_\text{br}$+O$_\text{top}$) surface -- or molecular way -- leading to (1$\times$1)(O$_\text{2dm}$) surface in 1 molecular layer (ML) concentration and/or (2$\times$1)(O$_\text{2dm}$+V$_\text{br}$) surface in 0.5 ML. The oxidation of the subreduced (2$\times$1)(O$_\text{br}$+V$_\text{br}$) surface induces (2$\times$1)(O$_\text{br}$+O$_\text{2dm}$) surface. Here, O$_\text{top}$ denotes oxygen atom adsorbed on the five-coordinated Sn atom (Sn$_\text{5c}$), and O$_\text{2dm}$ the oxygen dimer formed on the surface between Sn$_\text{5c}$ and the four-coordinated Sn (Sn$_\text{4c}$) atoms.

Concerning the oxidization of surface, we considered a variety of conceivable adsorption sites and configurations of adsorbate in the case of O$_2$ molecule ({\it i.e.}, O$_\text{2dm}$) on the surface. Through the calculation of binding energies, we picked out the most reasonable surfaces with the highest binding energy, which are just presented in Fig.~\ref{fig_surf1}. For instance, three different configurations for each O$_\text{2dm}$ containing surface with vertical, horizontal and tilted oxygen dimer were taken into account, and it turned out that the horizontal configurations had the highest binding energy. Wang {\it et al.}~\cite{Wang} also derived the same result but Habgood and Harrison~\cite{Habgood} reported that the tilted or twisted dimer is energetically favored. We also conducted a systematic calculation of lower coverage surfaces using $(4\times1)$ and $(2\times2)$ supercells, in which the binding energies decrease by less than 200 meV. Accordingly, we will consider $(1\times1)$ and $(2\times1)$ phases in the following, which are enough large to effectively establish the phase diagram of SnO$_2$(110) surface.

\begin{table}[!ht]
\begin{center}
\footnotesize
\caption{\label{tab_bindene}Oxygen binding energies on SnO$_2$(110) surfaces in the unit of eV/atom.}
\begin{tabular}{llc}
\hline
Species        & Phase  & Binding energy \\
\hline
O$_\text{br}$  & (1$\times$1)(O$_\text{br}$)                & $-2.79$ \\
               & (2$\times$1)(O$_\text{br}$+V$_\text{br}$)  & $-2.66$ \\
               & (2$\times$1)(O$_\text{br}$+O$_\text{2dm}$) & $-2.90$ \\
O$_\text{pl}$  & (1$\times$1)(O$_\text{br}$)                & $-2.80$ \\
               & (1$\times$1)(V$_\text{br}$)                & $-1.93$ \\
O$_\text{top}$ & (1$\times$1)(O$_\text{br}$+O$_\text{top}$) & $~~1.97$ \\
O$_\text{2dm}$ & (1$\times$1)(O$_\text{2dm}$)               & $-2.89$ \\
               & (2$\times$1)(O$_\text{2dm}$+V$_\text{br}$) & $-2.87$ \\
               & (2$\times$1)(O$_\text{br}$+O$_\text{2dm}$) & $-3.12$ \\
\hline
\end{tabular}
\end{center}
\end{table}
\normalsize
From the calculated binding energies, as listed in Table~\ref{tab_bindene}, we see that O$_\text{2dm}$ adsorption is stronger than O$_\text{br}$ adsorption; $-2.89$ eV in (1$\times$1)(O$_\text{2dm}$) vs. $-2.79$ eV in (1$\times$1)(O$_\text{br}$), $-2.66$ eV in (2$\times$1)(O$_\text{br}$+V$_\text{br}$) vs. $-2.87$ eV in (2$\times$1)(O$_\text{2dm}$+V$_\text{br}$), and $-3.12$ eV vs. $-2.90$ eV in (2$\times$1)(O$_\text{br}$+O$_\text{2dm}$). Meanwhile, O$_\text{br}$ and O$_\text{pl}$ have similar binding strength; $-2.79$ eV vs. $-2.80$ eV in (1$\times$1)(O$_\text{br}$) surface. On the contrary to these species, O$_\text{top}$ can not be adsorbed exothermically as they have the positive binding energy. On the other hand, when decreasing the coverage from 1 ML to 0.5 ML, the binding of oxygen to the surface becomes weaker; from $-2.79$ eV in (1$\times$1)(O$_\text{br}$) to $-2.66$ eV in (2$\times$1)(O$_\text{br}$+V$_\text{br}$), and from $-2.89$ eV in (1$\times$1)(O$_\text{2dm}$) to $-2.87$ eV in (2$\times$1)(O$_\text{2dm}$+V$_\text{br}$).

\begin{figure}[!ht]
\begin{center}
\includegraphics[clip=true,scale=0.07]{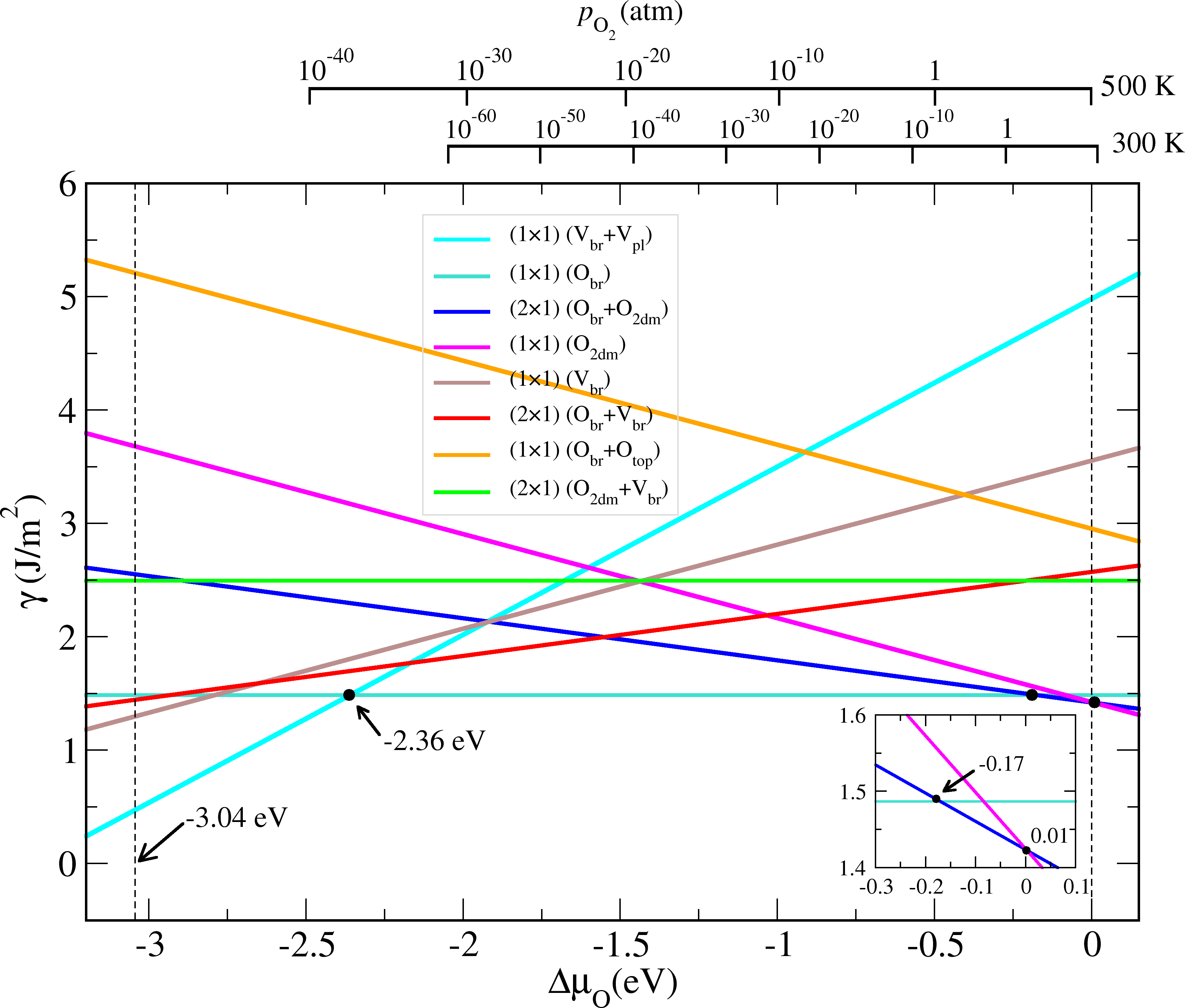}
\end{center}
\caption{\label{fig_phase1}(Color online) Surface Gibbs free energies of eight SnO$_2$(110) surface models suggested in this work and shown in Fig.~\ref{fig_surf1} as functions of oxygen chemical potential. Two dotted vertical lines indicate the lower and upper limit of oxygen chemical potential, where oxygen-poor limit is $1/2\Delta G^f_{\text{SnO}_2}(0, 0)=-3.04$ eV and oxygen-rich limit is $\Delta\mu_\text{O}=0.0$ eV. The pressure scales at fixed temperatures of $T=300$ K and $T=500$ K are presented in the top. Inset shows the intersections around oxygen-rich limit.}
\end{figure}

Then, we have calculated the Gibbs free energies of these eight surfaces by evaluating Eqs.~(\ref{ene_surf}) and (\ref{surf_ene4}) with $N_\text{NO}=0$. In Fig.~\ref{fig_phase1}, the calculated phase diagram of SnO$_2$(110) surface as a function of the oxygen chemical potential is presented. In the top of the figure, the pressure scales are marked at fixed temperatures of $T=300$ K and $T=500$ K, which are typical operating temperatures of SnO$_2$-based gas sensor. Note that the formation energy of stoichiometric (1$\times$1)(O$_\text{br}$) surface was calculated to be 1.49 J/m$^2$ with PBEsol+XDM method, which is higher than the value of 1.21 J/m$^2$ obtained by FP-LAPW method with PBE-GGA functional~\cite{Batzill05}. 

At lower oxygen chemical potentials from the oxygen-poor limit condition, which is $-3.04$ eV estimated from the bulk SnO$_2$ formation energy $-6.08$ eV, the fully reduced (1$\times$1)(V$_\text{br}$+V$_\text{pl}$) surface was estimated to be the most stable. Meanwhile, the stoichiometric (1$\times$1)(O$_\text{br}$) surface becomes energetically favored at higher oxygen chemical potentials from the value of $-2.36$ eV, which is in good agreement with other DFT result of $-2.4$ eV~\cite{Batzill04,Bergermayer,Agoston}. What is new in this work compared with other DFT calculations is that from $\Delta\mu_\text{O}=-0.17$ eV the stoichiometric (1$\times$1)(O$_\text{br}$) surface becomes less stable than the oxidized (2$\times$1)(O$_\text{br}$+O$_\text{2dm}$) surface. Also, another oxidized surface (1$\times$1)(O$_\text{2dm}$) has the lowest surface free energy over $\Delta\mu_\text{O}\approx0.01$ eV, but we will not consider this surface any more due to its being beyond the range (the oxygen-rich limit 0.0 eV). It is worthy noting here that the difference of Gibbs free energies between (1$\times$1)(O$_\text{br}$) and (2$\times$1)(O$_\text{br}$+O$_\text{2dm}$) is relatively small, {\it i.e.,} $\sim$0.06 J/m$^2$, but sufficiently higher than the numerical precision of surface energy calculation in this work, {\it i.e.,} 0.01 J/m$^2$. Furthermore, we should put special emphasis on the importance of vdW correction, since the intersection point without vdW correction (PBEsol only) was calculated to be $\sim$0.07 eV that is beyond the oxygen-rich limit 0.0 eV.

To sum up, the most stable SnO$_2$(110) surfaces according to the range of oxygen chemical potential are the fully reduced (1$\times$1)(V$_\text{br}$+V$_\text{pl}$) in the range of $(-3.04$, $-2.36)$ eV, the stoichiometric (1$\times$1)(O$_\text{br}$) in the range of $(-2.36$, $-0.17)$ eV, and the oxidized (2$\times$1)(O$_\text{br}$+O$_\text{2dm}$) in the range of $(-0.17, 0.0)$ eV. This result is reasonably consistent with the previous DFT works~\cite{Batzill04,Bergermayer,Agoston,Wang} and the general insight of surface stability.

\begin{figure}[!ht]
\begin{center}
\includegraphics[clip=true,scale=0.19]{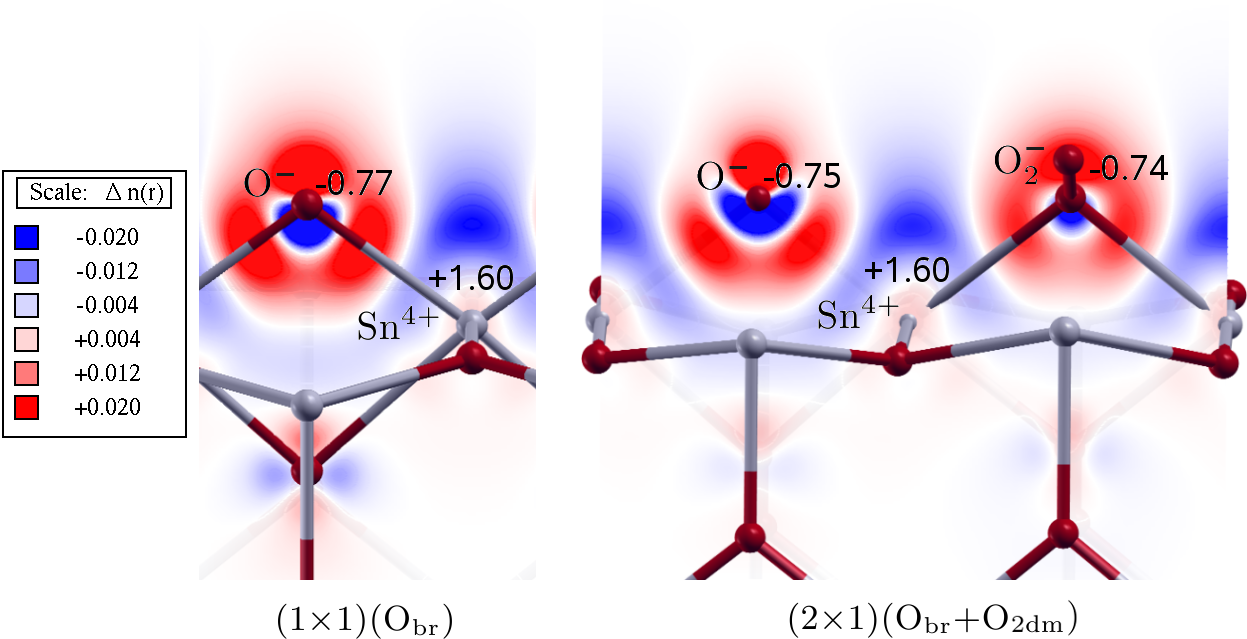}
\end{center}
\caption{\label{fig_dendiffo}(Color online) Electronic charge density difference in the formation of (1$\times$1)(O$_\text{br}$) and (2$\times$1)(O$_\text{br}$+O$_\text{2dm}$) surfaces. L\"{o}wdin charges of O$_\text{br}$, O$_\text{2dm}$, and Sn$_\text{6c}$ ions are presented. }
\end{figure}

Lastly in this subsection, we present the analysis of electronic charge transferring and L\"{o}wdin charges of ions in the formation of (1$\times$1)(O$_\text{br}$) and (2$\times$1)(O$_\text{br}$+O$_\text{2dm}$) surfaces. In Fig.~\ref{fig_dendiffo}, it is clearly shown that the electrons are transferred from Sn$_\text{6c}$ atom that becomes Sn$^{4+}$ ion to O$_\text{br}$ and/or O$_\text{2dm}$, which become O$^-$ and O$_2^-$ ions respectively. The calculated L\"{o}wdin charges of Sn$_\text{6c}$, O$_\text{br}$ and O$_\text{2dm}$ are 1.60 (this is almost same to that in the bulk), $-0.75$ and $-0.74$.

\subsection{\label{subsec_NOxadsorption}NO$_x$ adsorption on SnO$_2$(110) surface}
To commence a study of NO adsorption onto SnO$_2$(110) surface, we have made a modeling of NO-adsorbed surfaces, paying special attention on the location of possible surface adsorption sites and the geometry of adsorbate on the surface. Although it turned out in the previous subsection (\ref{subsec_o2surf}) that three surface configurations [(1$\times$1)(V$_\text{br}$+V$_\text{pl}$), (1$\times$1)(O$_\text{br}$), (2$\times$1)(O$_\text{br}$+O$_\text{2dm}$)] are favorable in energetics among eight different surfaces in the O$_2$ environment, we will use all the surface models considered there as the substrate surfaces of NO adsorption, except (1$\times$1)(O$_\text{br}$+O$_\text{top}$) surface that has positive oxygen binding energy. Therefore, seven different SnO$_2$(110) surfaces and additionally ($2\times1$)(2O$_\text{br}$) surface to take account of cell size effect were considered as the substrate surfaces hereafter.

\begin{figure*}[!ht]
\begin{center}
\includegraphics[clip=true,scale=0.27]{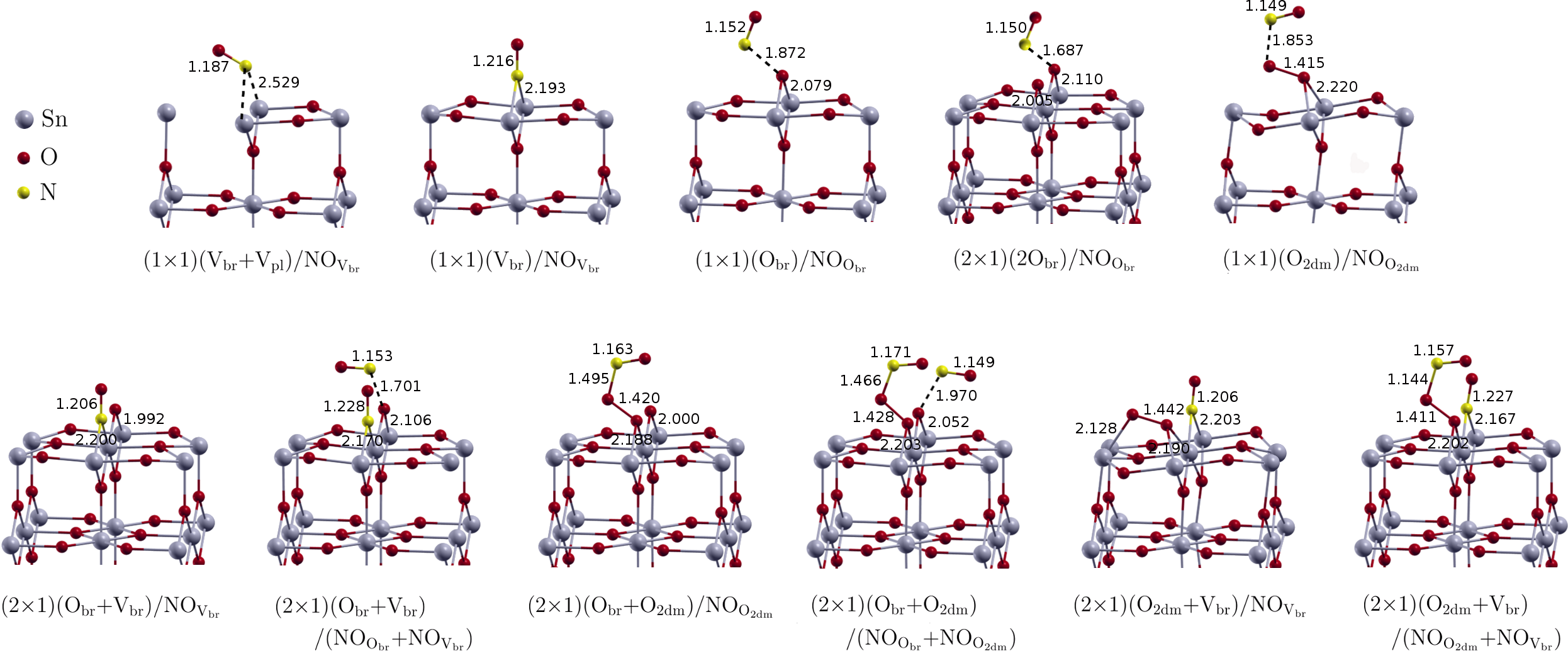}
\end{center}
\caption{\label{fig_surf2}(Color online) NO-adsorbed SnO$_2$(110) surfaces, selected as those with the largest adsorption energy among different adsorption configurations per each substrate surface. Atomic relaxation were performed with PBEsol+XDM method.}
\end{figure*}

For the possible adsorption sites of NO molecule, we can consider a variety number of sites such as bridging oxygen atoms (O$_\text{br}$), oxygen vacancy sites (V$_\text{br}$, V$_\text{pl}$), unsaturated Sn atoms (Sn$_\text{4c}$, Sn$_\text{5c}$) and oxygen dimer (O$_\text{2dm}$). Moreover, the possibility of two NO molecules adsorption for the cases of ($2\times1$) surfaces, being the coverage to be 1.0 ML, should not be missed. With respect to the geometry of adsorbed NO molecule, the direction of N-O bond axis to the substrate surface was considered: normal and parallel. For all such possible configurations of NO-adsorbed SnO$_2$(110) surface, we performed atomic relaxations and determined the adsorption energy of NO molecule to the surface using the following equation,
\begin{equation}
\label{ene_ads}
E_\text{ad}=\frac{1}{N_\text{mol}}[E_\text{surf+mol}-(E_\text{surf}+N_\text{mol}E_\text{mol})],
\end{equation}
where $E_\text{surf+mol}$ and $E_\text{surf}$ are the total energies of the surfaces with and without adsorbed molecule, and $N_\text{mol}$ and $E_\text{mol}$ are the number of adsorbed molecule and the total energy of isolated molecule, respectively. For the sake of simplicity, the surface with the largest adsorption energy among different configurations for each substrate was selected, being eleven different configurations, and presented in Fig.~\ref{fig_surf2}. The adsorption energies of these selected NO-adsorbed surfaces are listed in Table~\ref{tab_ads}. Here we adopted a notation for NO-adsorbed surfaces like substrate/NO$_\text{adsorption-site}$.
\begin{table}[!ht]
\begin{center}
\caption{\label{tab_ads}Adsorption energy (eV/molecule) of NO molecule onto the SnO$_2$(110) surfaces. The substrate surfaces are those presented in Fig.~\ref{fig_surf1}.}
\begin{tabular}{lcc}
\hline
NO-adsorbed surface & Coverage & $E_\text{ad}$ \\
\hline
(1$\times$1)(V$_{\text{br}}$+V$_{\text{pl}}$)/NO$_{\text{V}_{\text{br}}}$ & 1 & $-0.74$ \\
(1$\times$1)(V$_{\text{br}}$)/NO$_{\text{V}_{\text{br}}}$ & 1 & $-1.28$ \\
(1$\times$1)(O$_{\text{br}}$)/NO$_{\text{O}_{\text{br}}}$ & 1 & $-0.98$ \\
(2$\times$1)(2O$_{\text{br}}$)/NO$_{\text{O}_{\text{br}}}$ & 0.5 & $-1.17$ \\
(1$\times$1)(O$_{\text{2dm}}$)/NO$_{\text{O}_{\text{2dm}}}$ & 1 & $-0.87$ \\
(2$\times$1)(O$_{\text{br}}$+V$_{\text{br}}$)/NO$_{\text{V}_{\text{br}}}$ & 0.5 & $-1.57$ \\
(2$\times$1)(O$_{\text{br}}$+V$_{\text{br}}$)/(NO$_{\text{O}_{\text{br}}}$+NO$_{\text{V}_{\text{br}}}$) & 1 & $-1.11$ \\
(2$\times$1)(O$_{\text{br}}$+O$_{\text{2dm}}$)/NO$_{\text{O}_{\text{2dm}}}$ & 0.5 & $-1.20$ \\
(2$\times$1)(O$_{\text{br}}$+O$_{\text{2dm}}$)/(NO$_{\text{O}_{\text{br}}}$+NO$_{\text{O}_{\text{2dm}}}$) & 1 & $-0.90$ \\
(2$\times$1)(O$_{\text{2dm}}$+V$_{\text{br}}$)/NO$_{\text{V}_{\text{br}}}$ & 0.5 & $-1.29$ \\
(2$\times$1)(O$_{\text{2dm}}$+V$_{\text{br}}$)/(NO$_{\text{O}_{\text{2dm}}}$+NO$_{\text{V}_{\text{br}}}$) & 1 & $-1.06$ \\
\hline
\end{tabular}
\end{center}
\end{table}
\normalsize

When approaching NO molecule to $(1\times1)$(V$_\text{br}$+V$_\text{pl}$) surface, it is adsorbed at the V$_\text{br}$ site with an inclined geometry and an adsorption energy of $-0.74$ eV, then the substrate surface becomes NO-adsorbed surface denoted $(1\times1)$(V$_\text{br}$+V$_\text{pl}$)/NO$_{\text{V}_\text{br}}$. To the $(1\times1)$(V$_\text{br}$) substrate, the same adsorption site but normal geometry with an adsorption energy of $-1.28$ eV was identified. For the cases of $(1\times1)$(O$_\text{br}$) and $(2\times1)$(2O$_\text{br}$) substrates, NO molecules bond with the bridging oxygen atoms with inclined geometries and adsorption energies of $-0.98$ eV and $-1.17$ eV respectively. Meanwhile, in the case of $(1\times1)$(O$_\text{2dm}$) surface, the oxygen dimer O$_\text{2dm}$ was found to be the most preferable adsorption site with the adsorption energy of $-0.87$ eV. For one molecule adsorption on $(2\times1)$ surfaces, the bridging vacancy sites V$_\text{br}$ were confirmed to be the most natural adsorption sites, followed by the oxygen dimer O$_\text{2dm}$ and the bridging oxygen atom O$_\text{br}$, by inspecting $(2\times1)$(O$_\text{br}$+V$_\text{br}$)/NO$_{\text{V}_\text{br}}$, $(2\times1)$(O$_\text{br}$+O$_\text{2dm}$)/NO$_{\text{O}_\text{2dm}}$, and $(2\times1)$(O$_\text{2dm}$+V$_\text{br}$)/NO$_{\text{V}_\text{br}}$ surfaces. We note that other adsorption sites and geometries for each substrate were proved to be energetically less favorable than those shown in Fig.~\ref{fig_surf2}.

From Table~\ref{tab_ads}, we can assess the impact of adsorbate coverage. According to our calculation, an increase of coverage leads to a decrease of adsorption energy, and the adsorption energy when two NO molecules are adsorbed on (2$\times$1) surface is smaller than the sum of individual one-molecule adsorptions. This might be due to the repulsive interaction between adjacent NO molecules adsorbed on the surface. In the case of stoichiometric surface, for example, the adsorption energy at 1 ML coverage (one NO molecule on $(1\times1)$(O$_\text{br}$) substrate) is $-0.98$ eV versus $-1.17$ eV at 0.5 ML coverage (one molecule on $(2\times1)$(2O$_\text{br}$) substrate), and thus the NO-NO interaction energy is estimated to be $2\times(1.17-0.98)=0.38$ eV. For other surfaces, there happened also a increase of adsorption energy with a similar magnitude upon the change of coverage from 1 ML to 0.5 ML. The repulsive NO-NO interaction is likely to have an effect of restriction on the amount of adsorbate. At 1 ML coverage the NO-NO distance is $\sim$3.2 \AA~and at 0.5 ML coverage it is $\sim$6.4 \AA~that is enough large to ignore the NO-NO interaction. Therefore, with the coverages of 0.0 ML, 0.5 ML and 1.0 ML we can reasonably treat the adsorption and interaction of NO on the surface.

Regarding the geometry of NO adsorbate, it has been found that NO adsorbate at the V$_\text{br}$ site, except the $(1\times1)$(V$_\text{br}$+V$_\text{pl}$) surface, has a geometry of N-O bond axis normal to the surface, while at other sites they are inclined. For all the cases, the nitrogen atom rather than the oxygen atom of NO adsorbate always bonds with the surface atoms, resulting in a geometry of N-down orientation. On the other hand, the N-O bond length in the adsorbate at the V$_\text{br}$ site (1.187--1.228 \AA) slightly elongates compared with that in NO molecule (1.156 \AA) due to a stronger attraction of N atom toward surface Sn atoms, while at other sites they are more or less comparable (1.144--1.171 \AA), as shown in Fig.~\ref{fig_surf2}.

\begin{figure}[!ht]
\begin{center}
\includegraphics[clip=true,scale=0.17]{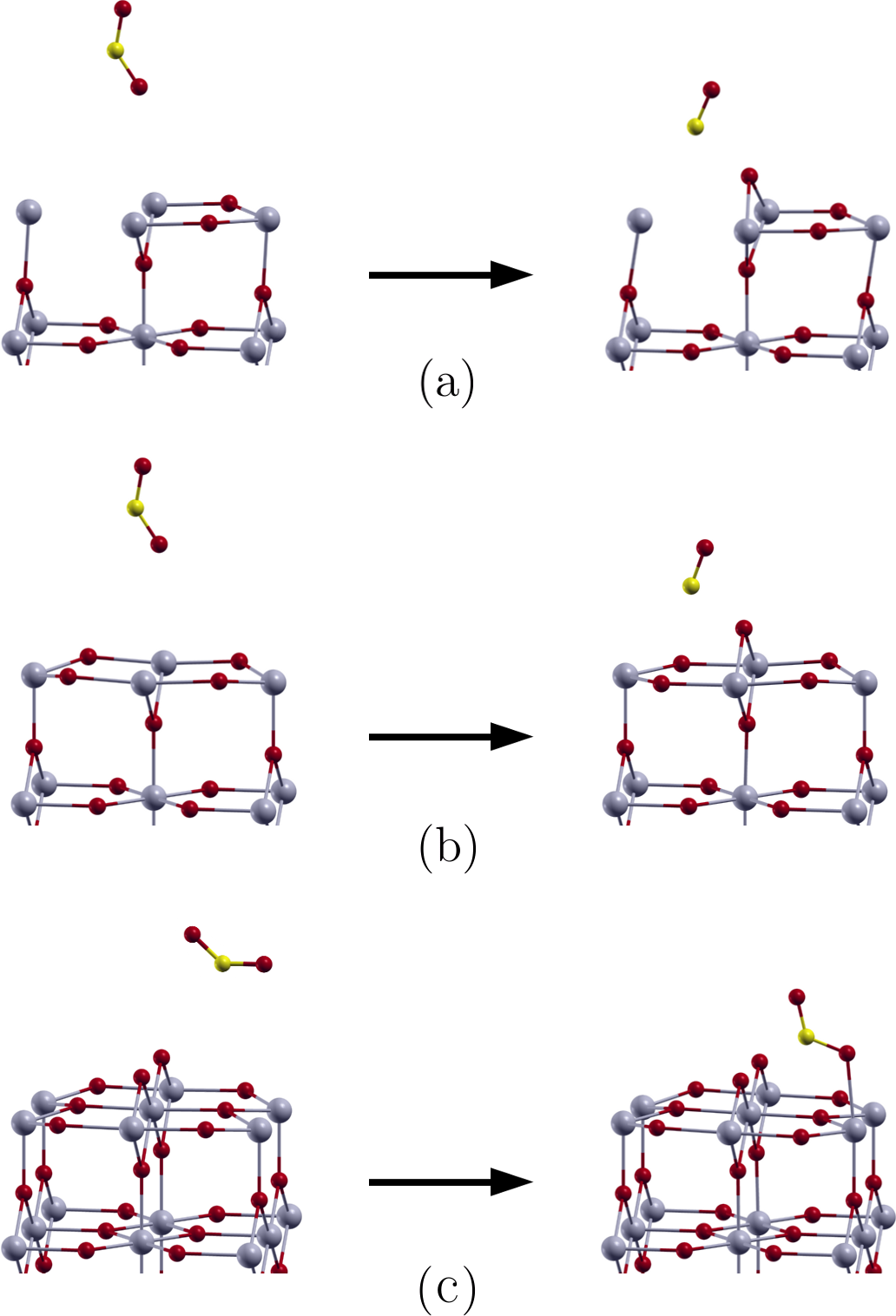}
\end{center}
\caption{\label{fig_NO2ads}(Color online) Adsorption of NO$_2$ on the SnO$_2$(110) containing typical oxygen vacancies and bridging oxygen atoms: (a) on $(1\times1)$(V$_\text{br}$+V$_\text{pl}$), (b) on $(1\times1)$(V$_\text{br}$), and (c) on $(2\times1)$(2O$_\text{br}$) surfaces.}
\end{figure}

We have also considered the adsorption of NO$_2$ molecule on the surfaces containing oxygen vacancies and the stoichiometric surface. As shown in Fig.~\ref{fig_NO2ads} (a) and (b), if NO$_2$ molecule approaches to the $(1\times1)$(V$_\text{br}$+V$_\text{pl}$) and $(1\times1)$(V$_\text{br}$) surfaces, it adsorbs on the V$_\text{br}$ sites. On the other hand, for the case of stoichiometric surface $(2\times1)$(2O$_\text{br}$), the Sn$_\text{5c}$ site was turned out to be thermodynamically favorable adsorption site. It is important noting that the geometry of NO$_2$ adsorbate shown in Fig.~\ref{fig_NO2ads} (b) is quite close to that of NO adsorbate in $(1\times1)$(O$_\text{br}$)/NO$_{\text{O}_\text{br}}$, while the surface configuration shown in Fig.~\ref{fig_NO2ads} (c) could be regarded as the final configuration for $(2\times1)$(O$_\text{br}$+O$_\text{2dm}$)/NO$_{\text{O}_\text{2dm}}$, as will be discussed in the following.

The results agree with previous work for NO$_x$ adsorption on SnO$_2$(110) surface~\cite{Prades07,Prades,Epifani08}, where preferred adsorption sites for NO and NO$_2$ are bridging oxygen atom and bridging-oxygen vacancies respectively~\cite{Prades07,Prades}, and Epifani {\it et al.}~\cite{Epifani08} also emphasize the role of bridging oxygen vacancies for NO$_2$ adsorption from an experimental and computational point of view. When compared with Al$_2$O$_3$ surface, NO adsorbate had also N-down orientation~\cite{Liu10}. In the case of CO adsorption on SnO$_2$(110) surface~\cite{Wang}, C-down orientations were mainly observed, being similar to NO adsorption.

\subsection{\label{subsec_phaseONO}Surface phase diagram in the O$_2$ and NO environment}
\begin{figure}[!ht]
\begin{center}
\includegraphics[clip=true,scale=0.54]{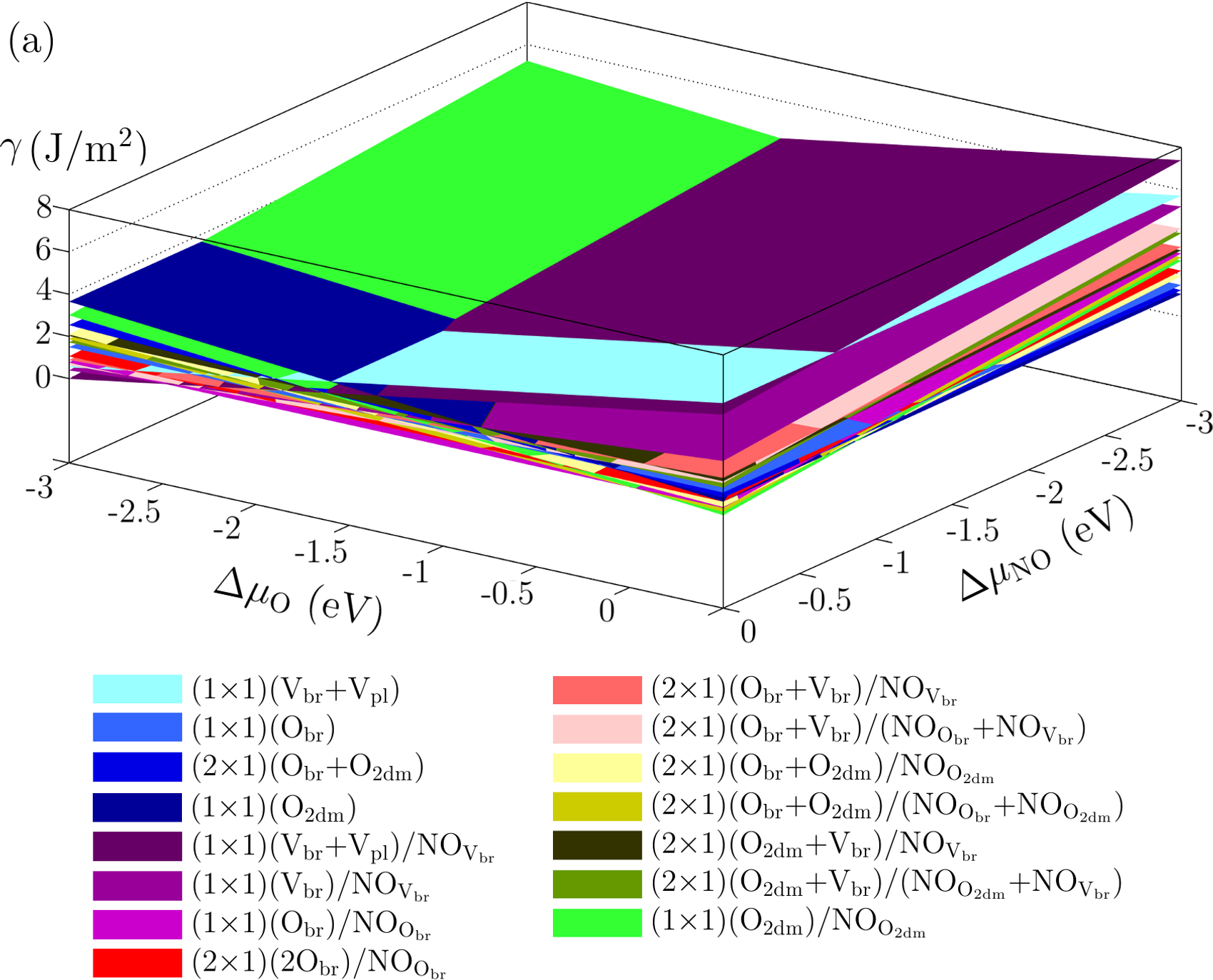} \\ \vspace{15pt}
\includegraphics[clip=true,scale=0.54]{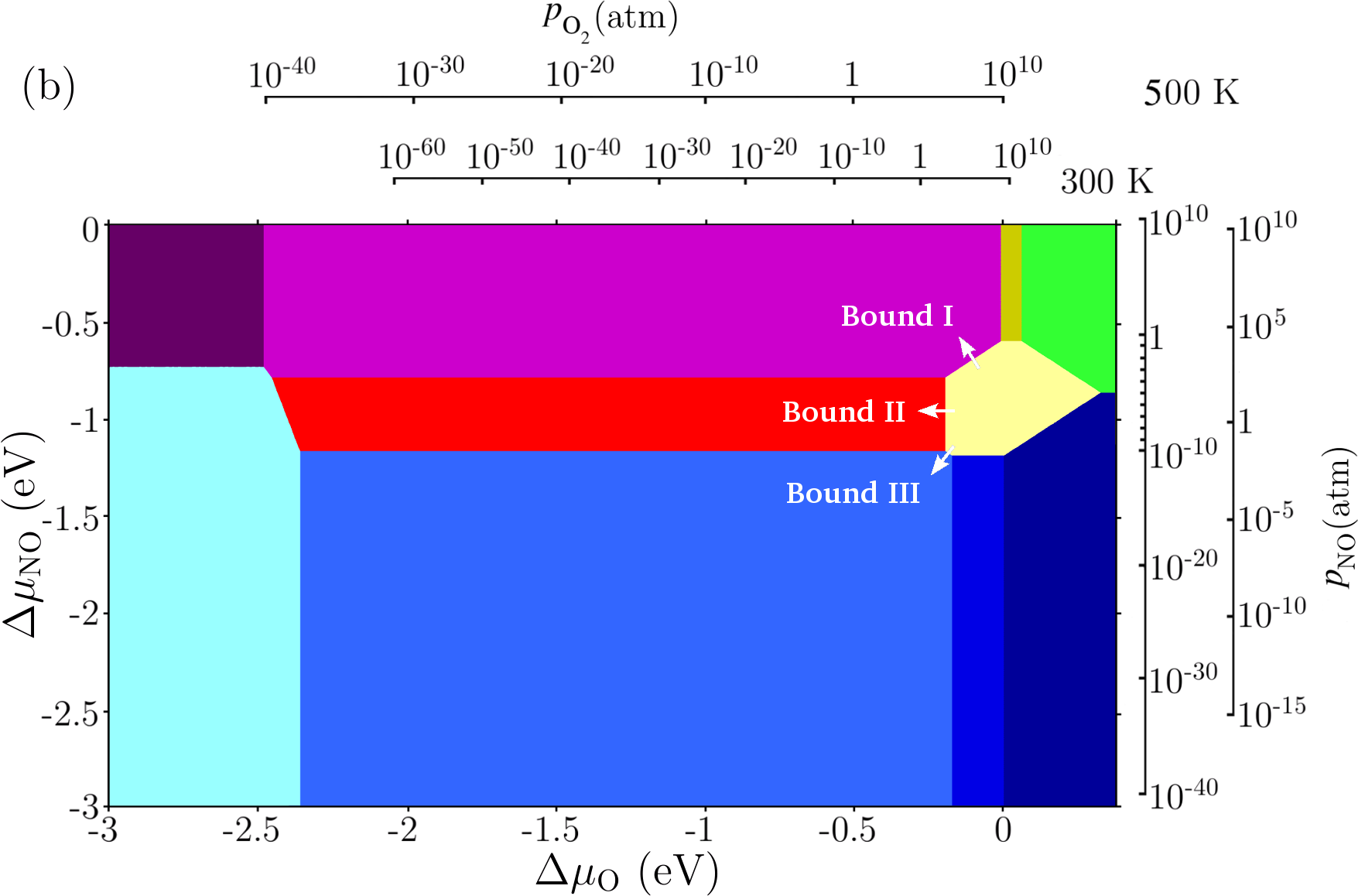}
\end{center}
\caption{\label{fig_phase2}(Color online) (a) Gibbs free energies of the eleven different surface phases shown in Fig.~\ref{fig_surf2} and (b) surface phase diagram of SnO$_2$(110) in ($\Delta\mu_\text{O}$, $\Delta\mu_\text{NO}$) space. At the top and right side in (b) the additional axes are shown to present the corresponding pressure scales at $T=300$ K and 500 K.}
\end{figure}

Since the most plausible phases of NO-adsorbed SnO$_2$(110) surface are in hand, we could determine the surface phase diagram of SnO$_2$(110) in the O$_2$ and NO environment by calculating Gibbs free energies of the eleven different surface phases, using Eq. (\ref{surf_ene4}). The results obtained with the PBEsol+XDM functional are plotted in Fig.~\ref{fig_phase2}, where (a) is a three dimensional graph showing the Gibbs free energies of the eleven NO-adsorbed surface phases according to both $\Delta\mu_\text{O}$ and $\Delta\mu_\text{NO}$, and (b) is just for surface phase diagram showing the most stable phases at any condition. In fact, the phase diagram of SnO$_2$(110) surface in contact with only O$_2$ gas (Fig.~\ref{fig_phase1}) is included in this 3D phase diagram; the information for the most stable phases in Fig.~\ref{fig_phase1} is exactly reflected in the line according to $\Delta\mu_\text{O}$ at NO-poor limit ($\Delta\mu_\text{NO}\approx-3.0$ eV) in Fig.~\ref{fig_phase2} (b).

From Fig.~\ref{fig_phase2} we can identify the most stable phases at $\Delta\mu_\text{NO}=-3.0$ eV (NO-poor limit) as $(1\times1)$(V$_\text{br}$+V$_\text{pl}$), $(1\times1)$(O$_\text{br}$) and $(2\times1)$(O$_\text{br}$+O$_\text{2dm}$), whose boundary points are $-2.36$ eV and $-0.17$ eV. These are agreed with the preceding result. When increasing the amount of NO gas in the environment, {\it i.e.}, increasing the partial pressure of NO gas, NO-adsorbed surfaces become thermodynamically favorable, which might be a natural process. Interestingly, the NO-adsorbed surfaces with the most stability are those formed by NO adsorption on the most stable surfaces at NO-poor limit condition. Within the possible scope of chemical potentials of oxygen ($-3.04$ eV $\leq\Delta\mu_\text{O}\leq0.0$ eV) and NO gases ($\Delta\mu_\text{NO}\leq0.0$ eV), they are $(1\times1)$(V$_\text{br}$+V$_\text{pl}$)/NO$_{\text{V}_\text{br}}$ (1 ML), $(1\times1)$(O$_\text{br}$)/NO$_{\text{O}_\text{br}}$ (1 ML) and $(2\times1)$(2O$_\text{br}$)/NO$_{\text{O}_\text{br}}$ (0.5 ML), and $(2\times1)$(O$_\text{br}$+O$_\text{2dm}$)/NO$_{\text{O}_\text{2dm}}$ (0.5 ML) and $(2\times1)$(O$_\text{br}$+O$_\text{2dm}$)/(NO$_{\text{O}_\text{br}}$+NO$_{\text{O}_\text{2dm}}$) (1 ML) surfaces. Similarly to the case of CO~\cite{Reuter03b}, $\Delta\mu_\text{NO}$ could in principle be varied down to $-\infty$, but we confine the range to $-3.0$ eV, at which the pure oxygen pre-adsorbed surfaces have already become most stable for any $\Delta\mu_\text{O}$, indicating that the NO content in the environment has become so low that NO can be no more stabilized at the surface. We also note that the last phase exists in negligible interval of $\Delta\mu_\text{O}$ around 0.0 eV, which will be ignored in the following discussion. We see in Fig.~\ref{fig_phase2} that the partial NO pressures for the surface phase transitions from $(1\times1)$(O$_\text{br}$) and $(2\times1)$(O$_\text{br}$+O$_\text{2dm}$) to the corresponding NO-adsorbed surfaces are $10^{-11}-10^{-10}$ atm at $T=300$ K and $10^{-2}-10^{-1}$ atm at 500 K.

Considering that the NO-adsorbed $(2\times1)$(O$_\text{br}$+O$_\text{2dm}$)/NO$_{\text{O}_\text{2dm}}$ surface can be thought as a metastable phase toward the NO$_2$-adsorbed surface shown in Fig.~\ref{fig_NO2ads} (c) due to their surface free energies, we have calculated the activation energy barrier to be negligibly low as 0.003 eV, indicating the almost spontaneous transition from the former phase to the latter phase. Therefore, it may safely be said that the $(2\times1)$(O$_\text{br}$+O$_\text{2dm}$)/NO$_{\text{O}_\text{2dm}}$ surface phase transforms readily into its boundary phases $(2\times1)$(2O$_\text{br}$)/NO$_{\text{O}_\text{br}}$ or $(1\times1)$(O$_\text{br}$)/NO$_{\text{O}_\text{br}}$ or $(1\times1)$(O$_\text{br}$) shown in Fig.~\ref{fig_phase2} (b). Regarding the chemical reactions, when increasing $\Delta\mu_\text{NO}$ (boundary between yellow and pink-colored parts: boundary I), the reaction can be written as follows,
\begin{multline}
\label{chemreact1}
(2\times1)(\text{O}_\text{br}+\text{O}_\text{2dm})/\text{NO}_{\text{O}_\text{2dm}}+2\text{NO}_\text{gas}\rightarrow 2(1\times1)(\text{O}_\text{br})/\text{NO}_{\text{O}_\text{br}}+\text{NO}_\text{2(ads)} \\
\leftrightarrow 2(1\times1)(\text{O}_\text{br})/\text{NO}_{\text{O}_\text{br}}+\text{NO}_\text{2(gas)}.
\end{multline}
When decreasing $\Delta\mu_\text{O}$, the reaction for the case of rich content of NO gas (yellow-red boundary: boundary II) is 
\begin{multline}
\label{chemreact2}
(2\times1)(\text{O}_\text{br}+\text{O}_\text{2dm})/\text{NO}_{\text{O}_\text{2dm}}+2\text{NO}_\text{gas}\rightarrow (2\times1)(2\text{O}_\text{br})/\text{NO}_{\text{O}_\text{br}}+\text{NO}_\text{2(ads)} \\
\leftrightarrow (2\times1)(2\text{O}_\text{br})/\text{NO}_{\text{O}_\text{br}}+\text{NO}_\text{2(gas)},
\end{multline}
while for the case of poor content of NO gas (yellow-blue boundary: boundary III) it is
\begin{multline}
\label{chemreact3}
(2\times1)(\text{O}_\text{br}+\text{O}_\text{2dm})/\text{NO}_{\text{O}_\text{2dm}}\rightarrow
(2\times1)(2\text{O}_\text{br})+\text{NO}_\text{2(ads)} \\
\leftrightarrow (2\times1)(2\text{O}_\text{br})+\text{NO}_\text{2(gas)}.
\end{multline}

In the above reactions, since the product surface phases are more stable than the reactant phases and the chemical potential of NO is much larger than NO$_2$, the forward reactions are occurred to create NO$_2$ gas. As mentioned above, the potential barriers for these reactions are negligibly low ($\sim$0.003 eV) and thus the oxidation of NO can be occurred rapidly under the catalysis of SnO$_2$ surface. If the content of created NO$_2$ gas attains to some extent and thus the chemical potential of NO$_2$ increases, the NO$_2$-adsorbed surface arrives at the thermodynamic equilibrium state. As a result, there exist both NO and NO$_2$ gases in the system, and therefore, the change of surface conductance by NO adsorption on the $(2\times1)(\text{O}_\text{br}+\text{O}_\text{2dm})$ surface may be further affected by NO$_2$ adsorption. On the adsorption of NO$_2$ on the stoichiometric SnO$_2$(110) surface, NO$_2$ molecule received a quite small charge of $0.04e$, indicating that the characteristics of conductance change during the surface phase transformations at boundary I and II due to NO adsorption on the $(2\times1)(\text{O}_\text{br}+\text{O}_\text{2dm})$ surface is similar to the case of NO adsorption on the $(1\times1)(\text{O}_\text{br})$ surface ({\it i.e.}, charge transferring to the surface$\rightarrow$decrease of potential barrier$\rightarrow$increase of conductance). On the other hand, in the case of phase transformation at boundary III, since the heights of potential barriers formed by O$_\text{br}$ are almost identical in both phases and the transition of O$_\text{2dm}$ to O$_\text{br}$ causes an additional charge transferring of $0.04e$ to the surface, no change occurred, considering the cancellation effect by NO$_2$ adsorption. In overall, NO adsorption on the $(2\times1)(\text{O}_\text{br}+\text{O}_\text{2dm})$ surface at oxygen-rich condition leads to the increase of surface conductance.

\subsection{\label{subsec_mechanism}NO$_x$ gas sensing mechanism}
\begin{figure}[!ht]
\begin{center}
\includegraphics[clip=true,scale=0.16]{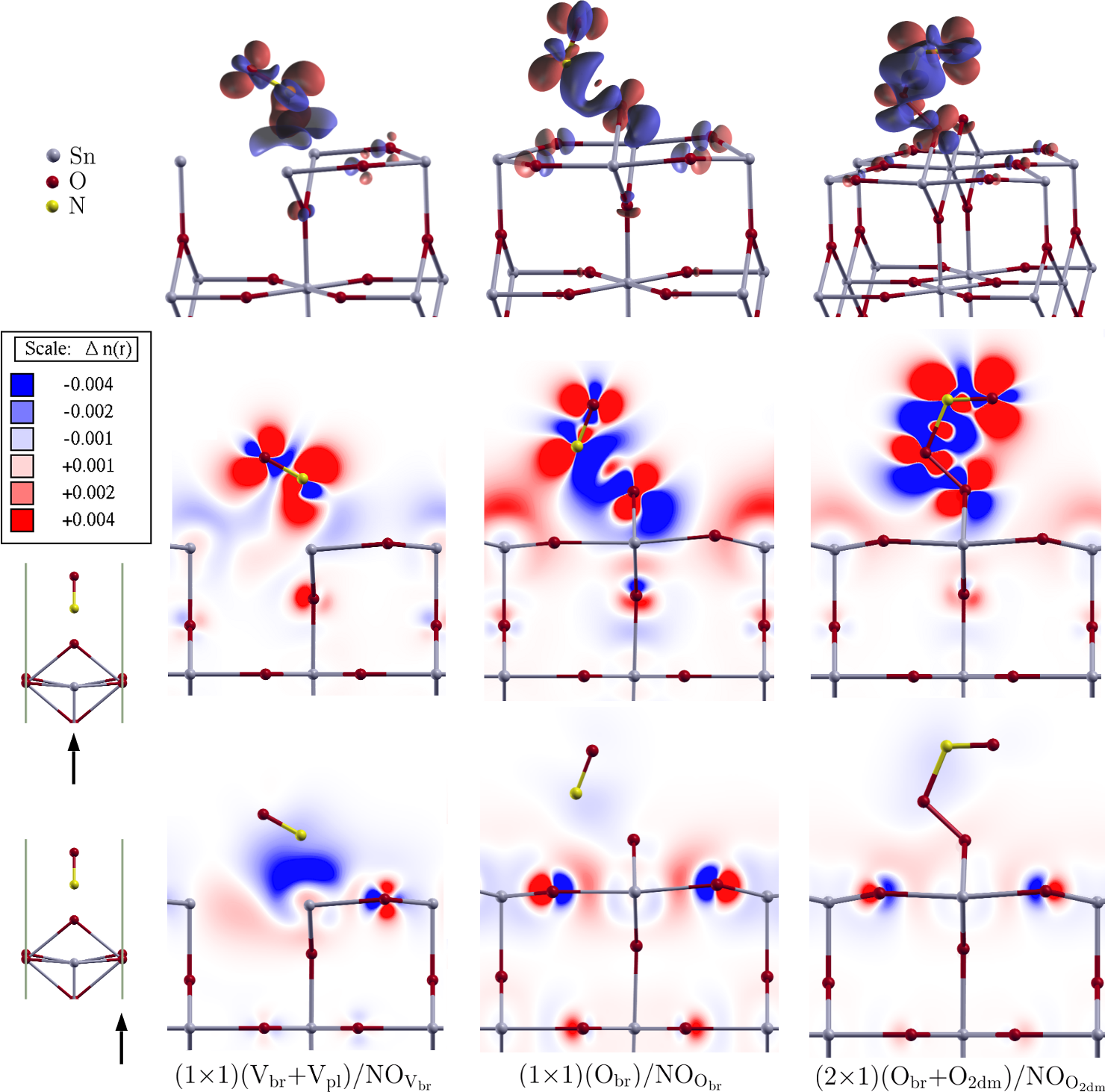}
\end{center}
\caption{\label{fig_den}(Color online) Electronic charge density difference in the event of NO adsorption on the most stable SnO$_2$(110) surfaces. Upper panel shows the 3D isosurfaces evaluated at the value of $\pm$0.004 $|e|$/\AA$^3$, where red (blue) color is for positive (negative) value indicating electron accumulation (depletion), and middle and lower panels show the 2D isoline pictures on the planes pointed by black arrows.}
\end{figure}

To uncover the sensing mechanism of SnO$_2$ to NO$_x$ gases, we consider in detail the electronic charge transferring in the NO-adsorbed SnO$_2$(110) surface phases. Fig.~\ref{fig_den} depicts the electronic charge density difference in the event of NO adsorption on the SnO$_2$(110) surfaces that are determined to be most stable from the surface phase diagram. From a careful analysis of L\"{o}wdin charges of atoms, we can estimate the electron transferring more quantitatively.

When a NO molecule was adsorbed on the $(1\times1)$(V$_\text{br}$+V$_\text{pl}$) surface at the oxygen poor limit, resulting in the formation of NO-adsorbed $(1\times1)$(V$_\text{br}$+V$_\text{pl}$)/NO$_{\text{V}_\text{br}}$ surface, NO receives a charge of $0.17e$ from the surface, resulting in the decrease of surface charge density and thus the increase of Schottky barrier leading to the decrease of surface charge conductivity. On the contrary, if NO was adsorbed on the $(1\times1)$(O$_\text{br}$) surface, NO lost a charge of $0.17e$ and the bridging oxygen atom O$_\text{br}$ also lost a charge of $0.12e$, indicating that total charge of $0.29e$ was transferred to the surface due to the indirect interaction between NO and the surface through O$_\text{br}$. In consequence, the concentration of surface charge carrier increases, the Schottky barrier remarkably decreases, and surface conductance increases. In addition, the generation of NO$^+-$O$^-_\text{br}$ polarization with opposite direction as well as the decrease of  O$^-_\text{br}-$surface polarization causes the decrease of potential barrier. In the case of NO adsorption on the $(2\times1)$(O$_\text{br}$+O$_\text{2dm}$) surface, the charges of NO and O$_\text{2dm}$ decrease by $0.24e$ and $0.01e$, but O$_\text{br}$'s charge increases by $0.05e$, indicating the similar charge transferring to the surface by indirect interaction between NO and surface through oxygen dimer O$_\text{2dm}$.

\begin{figure}[!ht]
\begin{center}
\includegraphics[clip=true,scale=0.1]{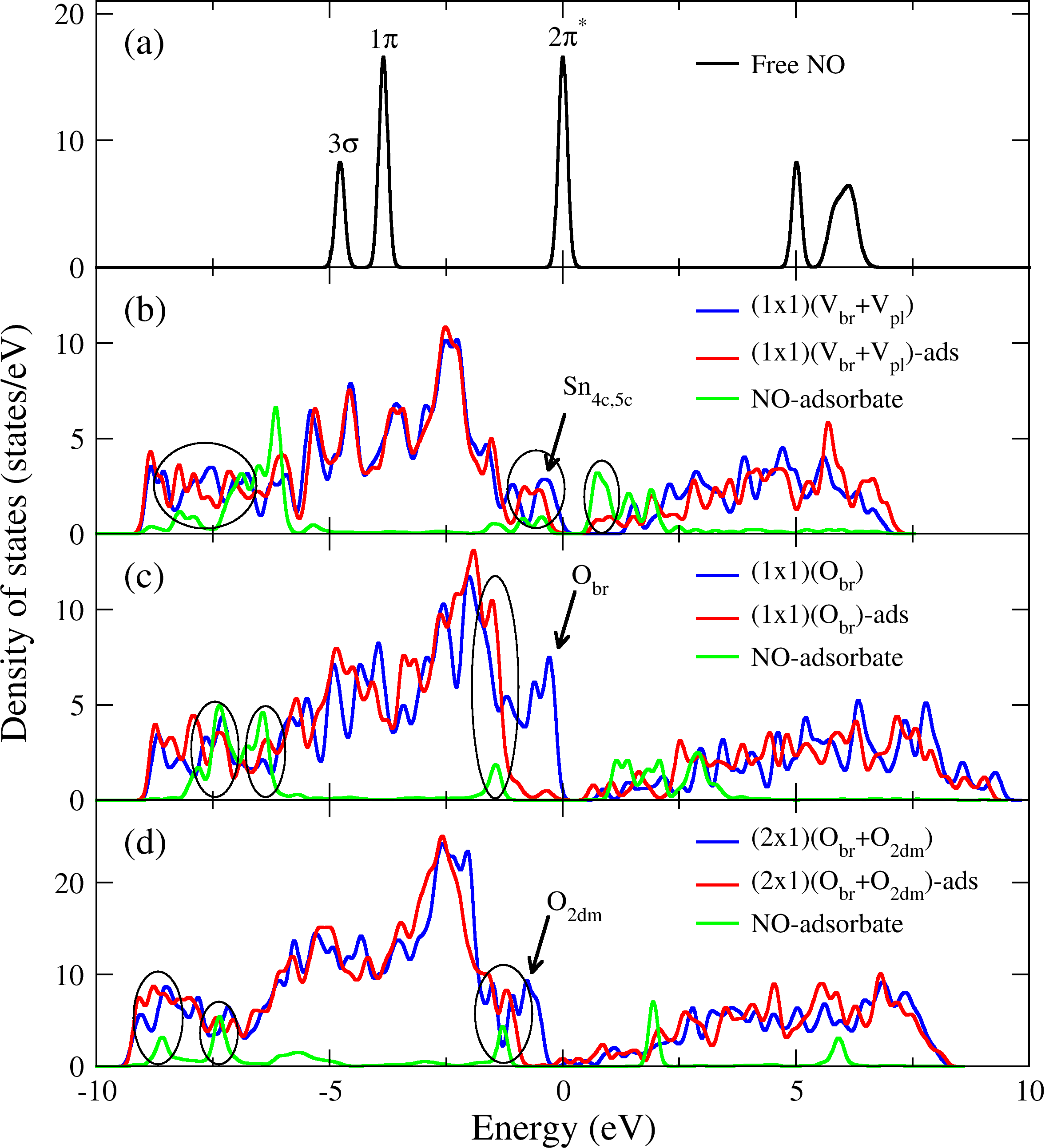}
\end{center}
\caption{\label{fig_dos}(Color online) Density of states of (a) free NO molecule, (b) $(1\times1)$(V$_\text{br}$+V$_\text{pl}$) surfaces before and after NO adsorption, (c) $(1\times1)$(O$_\text{br}$) surfaces and $(2\times1)$(O$_\text{br}$+O$_\text{2dm}$) surfaces. Fermi energy of each surface before NO adsorption is set to be 0 eV.}
\end{figure}

To clarify bonding characteristics of NO onto SnO$_2$(110) surfaces and the effects of NO adsorption on surface electronic structure, we calculated the density of states (DOS) for the most favorable SnO$_2$(110) surfaces before and after NO adsorption, and of free NO molecule, as shown in Fig.~\ref{fig_dos}. For (1$\times$1)(V$_\text{br}$+V$_\text{pl}$), surface valance states near the Fermi energy are mainly from 5s and 5p states of Sn$_\text{4c}$ and Sn$_\text{5c}$ atoms on the surface, while they are from O$_\text{br}$-2p states and O$_\text{2dm}$-2p states for (1$\times$1)(O$_\text{br}$) and (2$\times$1)(O$_\text{br}$+O$_\text{2dm}$) surfaces, respectively. In the case of (1$\times$1)(V$_\text{br}$+V$_\text{pl}$)/NO$_{\text{v}_\text{br}}$ surface, NO mainly interacts with Sn$_\text{4c}$ atom, causing hybridizations between NO-$1\pi$, $2\pi^*$ orbitals and Sn$_\text{4c}$-5s, 5p states, and the corresponding peaks are around $-7.5$ eV and $-0.8$ eV as shown in Fig.~\ref{fig_dos} (b). Meanwhile, there also occurred hybridizations between O$_\text{br}$-2p states and NO-$3\sigma$, $1\pi$ and $2\pi^*$ orbitals in the case of (1$\times$1)(O$_\text{br}$)/NO$_{\text{O}_\text{br}}$ surface, among which O$_\text{br}$-2p:NO-$3\sigma$ hybridization is a little weaker compared with the others. In this case, the resonances of peaks appear around $-6.4$ eV, $-7.5$ eV and $-1.4$ eV for hybridizations between O$_\text{br}$-2P states and NO-$3\sigma$, $1\pi$, $2\pi^*$ orbitals, respectively, as can be seen in Fig.~\ref{fig_dos} (c). Binding of O$_\text{2dm}$ and NO in the (2$\times$1)(O$_\text{br}$+O$_\text{2dm}$)/NO$_{\text{O}_\text{2dm}}$ surface is also based on 
hybridizations of O$_\text{2dm}$-2p states and NO-$3\sigma$, $1\pi$, $2\pi^*$ orbitals with 
localized peaks at about $-8.6$ eV, $-7.4$ eV and $-1.3$ eV in Fig.~\ref{fig_dos} (d).
The O$_\text{2dm}$-2p:NO $3\sigma$ hybridized states are approximately 2.2 eV lower than O$_\text{br}$-2p:NO-$3\sigma$ states, indicating the stronger combination of NO with O$_\text{2dm}$ on the (2$\times$1)(O$_\text{br}$+O$_\text{2dm}$)/NO$_{\text{O}_\text{2dm}}$ surface. NO adsorption on the SnO$_2$(110) surfaces in general induces a shift of the electronic states toward lower energy, and the magnitudes of shifts for (2$\times$1)(O$_\text{br}$+O$_\text{2dm}$) and (1$\times$1)(O$_\text{br}$) is more notable than for (1$\times$1)(V$_\text{br}$+V$_\text{pl}$), which is consistent with our exothermic NO adsorption energies of $-1.20$ eV, $-0.98$ eV and $-0.74$ eV on (2$\times$1)(O$_\text{br}$+O$_\text{2dm}$), (1$\times$1)(O$_\text{br}$) and (1$\times$1)(V$_\text{br}$+V$_\text{pl}$) surfaces, respectively. It is important to note that NO adsorption causes the narrowing of the band gap of
(1$\times$1)(V$_\text{br}$+V$_\text{pl}$), while not for both (1$\times$1)(O$_\text{br}$) and (2$\times$1)(O$_\text{br}$+O$_\text{2dm}$), due to the creation of new hybridized states between Sn$_\text{4c}$-5p states and unoccupied NO orbitals at the bottom of conduction band.

In the sensing mechanism of SnO$_2$ compared to other gas like CO and ethanol, it was described that CO adsorption on the surface causes the taking off the surface oxygen atoms like O$_\text{br}$ and thus reduction of the surface~\cite{Wang16,Wang,Yang}. Unlike this, it is not easy for NO to desorb O$_\text{br}$ but only possible to take off one oxygen atom of O$_\text{2dm}$ because of relatively weak reducibility of NO. Moreover, the characteristics of conductance variation is dependent on the adsorption site: V$_\text{br}$--decrease, O$_\text{br}$--increase, and O$_\text{2dm}$ almost constant. Therefore, the concentrations of environmental oxygen and oxygen vacancy are main tuning parameters to improve the sensibility of SnO$_2$ materials.

Since there could be a lot of defects like oxygen vacancy created by different factors (impurities, defects, etc.) accompanied with materials synthesis of practical SnO$_2$ sensors, all kinds of adsorptions of NO and NO$_2$ on different sites at any sensing condition occur with only different portions. The NO$_2$ gas generated by the reactions~\ref{chemreact1}--\ref{chemreact3} may be adsorbed on oxygen vacancies even a little, having an effect on conductance variation. As pointed out in Ref.~\cite{Epifani08}, at the condition of high oxygen partial pressure and high temperature, the sensitivity of SnO$_2$ to NO$_2$ is degraded due to decrease of oxygen vacancy concentration caused by filling of adsorbed oxygen into the vacancy site. According to the nudged elastic band calculation, the activation energy of transformation from $(2\times1)(\text{O}_\text{2dm}+\text{V}_\text{br})$ to $(2\times1)(2\text{O}_\text{br})$ was determined to be 0.60 eV.

\section{\label{sec_sum}Summary}
In the present work, we have investigated the SnO$_2$(110) surfaces in contact with oxygen and NO gases by means of {\it ab initio} atomistic thermodynamic method, aiming to find out the gas sensing mechanism for NO and NO$_2$ gases. We presented a detailed formalism for calculating the Gibbs free energies of NO-adsorbed SnO$_2$(110) surfaces, considering both O$_2$ and NO chemical potentials, on which the stability of surfaces depends strongly.

As preliminary stage for gas sensing, SnO$_2$(110) surfaces are pre-adsorbed by oxygen, forming the surface O$^-$ and O$^-_2$ ions. Using (1$\times$1) and (2$\times$1) surface unit cells, we have built the eight different surface slab models and then determined the most stable surfaces in contact with only oxygen gas. Being in good agreement with the previous works, the fully reduced surface containing the bridging and in-plane vacancies in the oxygen-poor condition ($-3.04$ eV $<\Delta\mu_\text{O}<-2.36$ eV), the fully oxidized surface containing the bridging oxygen and oxygen dimer in the oxygen-rich condition ($-0.17$ eV $<\Delta\mu_\text{O}<0.0$ eV), and the stoichiometric surface in between were thermodynamically most favorable. The creation of O$^-$ and O$^-_2$ ions was confirmed.

With the calculation of adsorption energy, we have identified the most preferable adsorption sites for NO molecule on the oxygen pre-adsorbed surfaces, as the bridging vacancy sites, the oxygen dimer and the bridging oxygen atoms. The geometry of N-down orientation was observed for all adsorbates. Using the selected most plausible NO-adsorbed surfaces, the surface phase diagram in ($\Delta\mu_\text{O}$, $\Delta\mu_\text{NO}$) space was determined. At the NO-rich condition, the most stable surfaces were those formed by NO adsorption on the most stable surfaces in only oxygen contact. In the excess of NO gas, the surface phase transitions occurred due to the interaction between NO and the NO-adsorbed surface, creating NO$_2$ molecule. The electronic charge density difference to estimate the charge transferring during the NO adsorption and density of states to describe the chemical bonding characteristics were calculated.

With respect to the gas sensing mechanism, NO adsorption on oxygen vacancy sites causes a decrease of conductance at oxygen-poor condition (rich of oxygen vacancies), while at oxygen-rich condition NO adsorption on bridging oxygen sites leads to an increase of conductance. The interaction between the adsorbed NO and oxygen atom on the surface readily leads to creation of NO$_2$, and the rise of NO$_2$ concentration has also effect on the conductance variation. To make sensing well for NO gas at the oxygen rich condition, therefore, the decrease of oxygen vacancy concentration as much as possible by careful control of synthesis and preprocess of the sample is necessarily indispensable to increase the conductance by removing the interference of NO$_2$ gas. On the contrary, when sensing for NO and NO$_2$ gases at the oxygen poor condition, the decrease of conductance leads to a good sensing.

\section*{\label{ack}Acknowledgments}
This work was supported partially from the Committee of Education, Democratic People's Republic of Korea, under the project entitled ``Strong correlation phenomena at superhard, superconducting and nano materials'' (Grant number 02-2014). The simulations have been carried out on the HP Blade System c7000 (HP BL460c) that is owned and managed by the Faculty of Materials Science, Kim Il Sung University.

\bibliographystyle{apsrev}
\bibliography{Reference}

\end{document}